\documentclass[3p]{elsarticle}
\usepackage{lineno}
\modulolinenumbers[5]
\journal{Acta Materialia}
\usepackage{graphicx,ulem}
\usepackage{xspace} 
\xspaceaddexceptions{)}
\xspaceaddexceptions{]}
\usepackage{textcomp}
\usepackage[]{lmodern}
\usepackage{siunitx} 
\usepackage{tabularx}
\usepackage{floatrow}
\usepackage[font={Large,bf},labelformat=simple]{subfig} 

\usepackage{color}
\usepackage{amsmath}
\usepackage{soul}
\usepackage{booktabs}
\biboptions{numbers,sort&compress}
\usepackage{listings}
\usepackage{paralist}
\usepackage{epstopdf}
\usepackage{setspace}
\usepackage[export]{adjustbox}
\usepackage[scaled=1.05]{gentium}
\usepackage{textcomp,fancyhdr}
\usepackage[plainpages=false]{hyperref}
\hypersetup{final} 
\usepackage[capitalise]{cleveref}
\newcommand{\sic}[1]{\SI{#1}{\celsius}\xspace}

\renewcommand{\eqref}[1]{Eq.~(\ref{#1})\xspace}
\let\subtef\subref
\renewcommand{\subref}[1]{\protect\subtef*{#1}}

\renewcommand{\bar}[1]{\mkern 1.25mu\overline{\mkern-1.25mu#1\mkern-1.25mu}\mkern 1.25mu}

\newcommand{\changes}[1] {\color{black} #1 \color{black}}

\newcommand{\rms}{sd}

\def\ep{\varepsilon^{\rm P}\xspace}
\def\ef{\varepsilon_{\rm fs}\xspace}
\def\erate{\dot{\varepsilon}\xspace}
\def\eprate{\dot{\varepsilon}\xspace}
\def\einf{\varepsilon^{\infty}\xspace}
\def\sinf{\sigma^{\infty}\xspace}
\def\eavg{{e}_{\rm f}\xspace}

\crefmultiformat{figure}%
{\edef\crefstripprefixinfo{#1}Figs.~#2#1#3}%
{ and~(#2\crefstripprefix{\crefstripprefixinfo}{#1}#3}%
{, #2\crefstripprefix{\crefstripprefixinfo}{#1}#3}%
{, and~#2\crefstripprefix{\crefstripprefixinfo}{#1}#3}

\begin{document}

\begin{frontmatter}

  \title{Creep failure of honeycombs made by rapid prototyping}

  \author[cam]{P.~E.~Seiler} \ead{pes34@cam.ac.uk}
  \author[cam]{H.~C.~Tankasala} \ead{hct30@cam.ac.uk}
  \author[cam]{N.~A.~Fleck\corref{cor}} \ead{naf1@eng.cam.ac.uk}
  
  \cortext[cor]{Corresponding author}
  
  \address[cam]{Department of Engineering, University of Cambridge,
    Cambridge CB2 1PZ, United Kingdom}

\begin{abstract}
  Additive manufacture and rapid prototyping are versatile methods for
  the generation of lattice materials for applications in the creep
  regime. However, these techniques introduce defects that can degrade
  the macroscopic creep strength. In the present study, the uniaxial
  tensile response of two-dimensional PMMA lattices is measured in the
  visco-plastic regime: tests are performed at \sic{100} which is
  slightly below the glass transition temperature $T_\mathrm{g}$ of
  PMMA. Both \textit{as-manufactured} defects (Plateau borders and
  strut thickness variation) and \textit{as-designed} defects (missing
  cell walls, solid inclusions, and randomly perturbed joints) are
  introduced. The dispersion in macroscopic strength is measured for
  relative densities in the range of 0.07 to 0.19. It is observed that
  initial failure of the lattice is diffuse in nature: struts fail at
  a number of uncorrelated locations, followed by the development of a
  single macroscopic crack transverse to the loading direction. In
  contrast, the same PMMA lattice fails in a correlated, brittle
  manner at room temperature.  An FE study is performed to gain
  insight into the diffuse failure mode and the role played by
  \textit{as-manufactured} defects, including the dispersion in
  tensile strength of individual struts of the lattice. A high damage
  tolerance to \textit{as-designed} defects is observed
  experimentally: there is negligible knock-down in strength due to
  the removal of cell walls or to the presence of solid
  inclusions. These findings aid the design and manufacture of damage
  tolerant lattices in the creep regime.
\end{abstract}


\end{frontmatter}

\section{Introduction}
\label{sec:introduction}
Foams and lattices are increasingly used in engineering applications
due to advances in additive manufacturing methods. \changes{A recent
  example is the use of micro-architectured polymeric soles of running
  shoes.} There is a need to develop tough, stiff, and lightweight
lattices, with high creep resistance for high temperature
applications, e.\,g.~heat exchangers, thermal insulation and catalytic
converters~\cite{Gibson1999}. Whilst the creep behaviour of foams has
been
reported~\cite{Andrews1999b,Oppenheimer2007,Mills1997,Deschanel2009,Zhu1999,Fan2012,Khan2017,Huang1991a,Huang2003,Goretta1990},
few studies detail the tensile response of rate-sensitive,
visco-plastic lattices~\cite{Oppenheimer2007,Chen2008b}. The present
study addresses this gap in the literature, especially in regard to
as-manufactured and as-designed imperfections and the implication of
these two classes of imperfection on the macroscopic response.  The
present study is companion to that of the recent
paper~\cite{Seiler2019a} on the brittle response of polymethyl
methacrylate (PMMA) lattices tested at room temperature. In
particular, the defect-sensitivity of creep strength is compared with
that of brittle strength as taken from~\cite{Seiler2019a}, for the
same geometries and same manufacturing technique.

\subsection{Hexagonal lattices}
\label{sec:brittle_lattices}
A regular two-dimensional (2D) hexagonal lattice is shown in
\cref{fig:PerfectLatticeSketch}. It comprises struts of length $\ell$
and in-plane thickness $t$ such that, for $t/\ell<0.2$, the relative
density $\bar{\rho}$ of the lattice is given by
\begin{equation}
  \label{eq:relative_density}
  \bar{\rho}=\frac{2}{\sqrt{3}} \frac{t}{\ell}
\end{equation}
as derived in Ref.~\cite{Gibson1999}. Under macroscopic uniaxial loading, the
hexagonal lattice is
bending-dominated~\cite{Gibson1999}. Consequently,
the macroscopic stiffness $E$ and the macroscopic strength
$\sigma^\infty_\mathrm{f}$ of a brittle hexagonal lattice scale with relative density $\bar{\rho}$
according to
\begin{equation}
  \label{eq:stiffnes_hex}
  E=\frac{3}{2}\bar{\rho}^3E_\mathrm{s} \mathrm{\hspace{0.5cm}and\hspace{0.5cm}}
  \sigma^\infty_\mathrm{f}=\frac{1}{3}\bar{\rho}^2\sigma_\mathrm{fs}
\end{equation}
in terms of the Young's modulus $E_\mathrm{s}$ and tensile strength
$\sigma_\mathrm{fs}$ of the parent solid. In the present study, we
select polymethyl methacrylate (PMMA) as the cell wall material of the
hexagonal lattice since it behaves in an elastic-brittle manner at
room temperature, but is visco-plastic at temperatures close to the
glass transition temperature, $T_\mathrm{g}$.

The presence of as-manufactured defects significantly degrade the
macroscopic tensile strength of elastic-brittle hexagonal
lattices~\cite{Seiler2019a}. For example, a
dispersion in strut thickness and in tensile strength of individual
struts within the lattice knocks down the macroscopic
strength. Successive strut failure occurs in the vicinity of a
previously failed strut, and this leads to a correlated failure
mode. Ultimately, a single macroscopic crack nucleates and advances
from one edge of the lattice. This failure mechanism was also predicted in the finite element (FE) simulations of elastic-brittle PMMA lattices by
assuming a local tensile strain failure criterion. Furthermore, both
the measured and predicted macroscopic strength of a lattice of finite
specimen geometry agree with the analytical predictions of
\citet{Gibson1999}. The strength of the brittle lattice is sensitive
to the presence of as-designed defects
in the form of missing cell walls\footnote{The removal of cell walls
  leads to a shift in behaviour from strength control to $K$-dominated
  behaviour at a small transition flaw size $a_\mathrm{T}$ on the
  order of 1 cell size.} and it is relatively
insensitive to the presence of filled cells and randomly misplaced joints. The aim of the
present study is to determine whether similar conclusions can be drawn
for a lattice deforming in the creep regime. For example, is the
degree of imperfection sensitivity in the creep regime significantly
less than that in the elastic-brittle regime?

\subsection{The creep response of foams and honeycombs}
\label{sec:effect_relative_density_creep_rate}
At elevated temperature $T$ close to $T_\mathrm{g}$ for polymers, or
$T$ exceeding 30\% of the absolute melting temperature $T_\mathrm{m}$
for metals and ceramics, foams undergo time-dependent plastic flow, or
\textit{creep}. \citet{Gibson1999} have considered the case of a foam
with cell walls that undergo power-law creep, such that the uniaxial
strain rate $\dot\varepsilon$ of the solid scales with the tensile
stress $\sigma$ according to

\begin{equation}
  \label{eq:power_law_creep}
  \dot\varepsilon = \dot\varepsilon_\mathrm{s} \left( \frac{\sigma}{\sigma_\mathrm{s}}\right)^{n_\mathrm{s}}
\end{equation}
in terms of the material constants $\dot\varepsilon_\mathrm{s}$,
$\sigma_\mathrm{s}$, and the creep exponent $n_\mathrm{s}$. The creep exponent of
the honeycombs or foams inherits its value from that of the parent
solid, as explained by~\citet{Boccaccini2014}. \citet{Gibson1999}
predict that the macroscopic creep rate $\dot\varepsilon^\infty$ of an
open-cell foam scales with the macroscopic tensile stress
$\sigma^\infty$ according to
\begin{equation}
  \label{eq:creep_rate_relDensity}
  \frac{\dot\varepsilon^\infty}{\dot\varepsilon_\mathrm{s}} =
  \frac{0.6}{(n_\mathrm{s}+2)} \left(
    \frac{1.7(2n_\mathrm{s}+1)}{n_\mathrm{s}}
    \frac{\sigma^\infty}{\sigma_\mathrm{s}}\right)^{n_\mathrm{s}}
  \left(\frac{1}{\bar{\rho}} \right)^{(3n_\mathrm{s}+1)/2}
\end{equation}
The dependence of creep rate upon relative density, as stated in
\cref{eq:creep_rate_relDensity}, has been confirmed experimentally for
open-cell metallic~\cite{Andrews1999b} and ceramic
foams~\cite{Goretta1990}. The creep parameters of polymeric foams also
inherit their values from those of the parent solid; for example,
\citet{Huang1991a} found that the creep activation energy of foamed
polystyrene equals that of the solid. The visco-elastic response of
open and closed-celled polymer foams has been studied in the
low-temperature
regime~\cite{Mills1997,Zhu1999,Huang1991a,Deschanel2009} whereas the
present study addresses the high-temperature visco-plastic, creep
response of polymer honeycombs at temperatures close to the glass
transition temperature, $T_\mathrm{g}$.

It is broadly accepted that the power-law creep response of honeycombs
and foams is sensitive to the presence of missing cell
walls~\cite{Su2014,Huang2003}, to a dispersion in cell wall
thickness~\cite{Han2018}, to the existence of randomly misplaced
joints~\cite{Han2018}, and Plateau
borders~\cite{Lin2005,Chen2008b}. The emphasis of the present study
differs from that of previous work on the creep of polymeric foams:
our aim is to explore the sensitivity of failure mechanisms (localised
versus diffuse), and associated strength and ductility, to the
presence of geometric and material imperfections.

\subsection{Scope of study}
\label{sec:scope}
The purpose of the current study is to examine experimentally the
deformation and fracture responses of two-dimensional (2D),
visco-plastic, hexagonal honeycombs made by rapid prototyping. The
honeycombs were cut from polymethyl methacrylate (PMMA) sheets and the
macroscopic stress versus strain response was measured for uniaxial
tension at fixed strain rate, in a temperature-controlled chamber at
\sic{100} which is close to the glass transition temperature
$T_\mathrm{g}$ of PMMA. The as-manufactured lattice geometry contains
Plateau borders and strut thickness variation. The magnitude of these
2 defects was measured by computer-assisted tomography (CT). Finite
element (FE) predictions of the macroscopic stress versus strain
response were made on the basis of (i) the measured geometry from the
CT-scans, (ii) the measured visco-plastic response of a single strut,
and (iii) the measured dispersion of strut ductility associated with
the manufacturing method. The notion of a transition flaw size is used
to quantify the sensitivity of lattice strength to defect
size. Irregular lattices were also created by the introduction of
as-designed defects, specifically a centre crack (due to missing cell
walls), solid inclusions in the form of filled cells, and randomly
perturbed joints. The knock-down in lattice strength due to each of
these defects was measured.

\section{Experimental investigation}
\label{sec:experimental_programme}
\begin{figure}[htbp]
  \centering \setlength{\labelsep}{-0.25cm}
  \sidesubfloat[]{\includegraphics[width=0.35\textwidth]{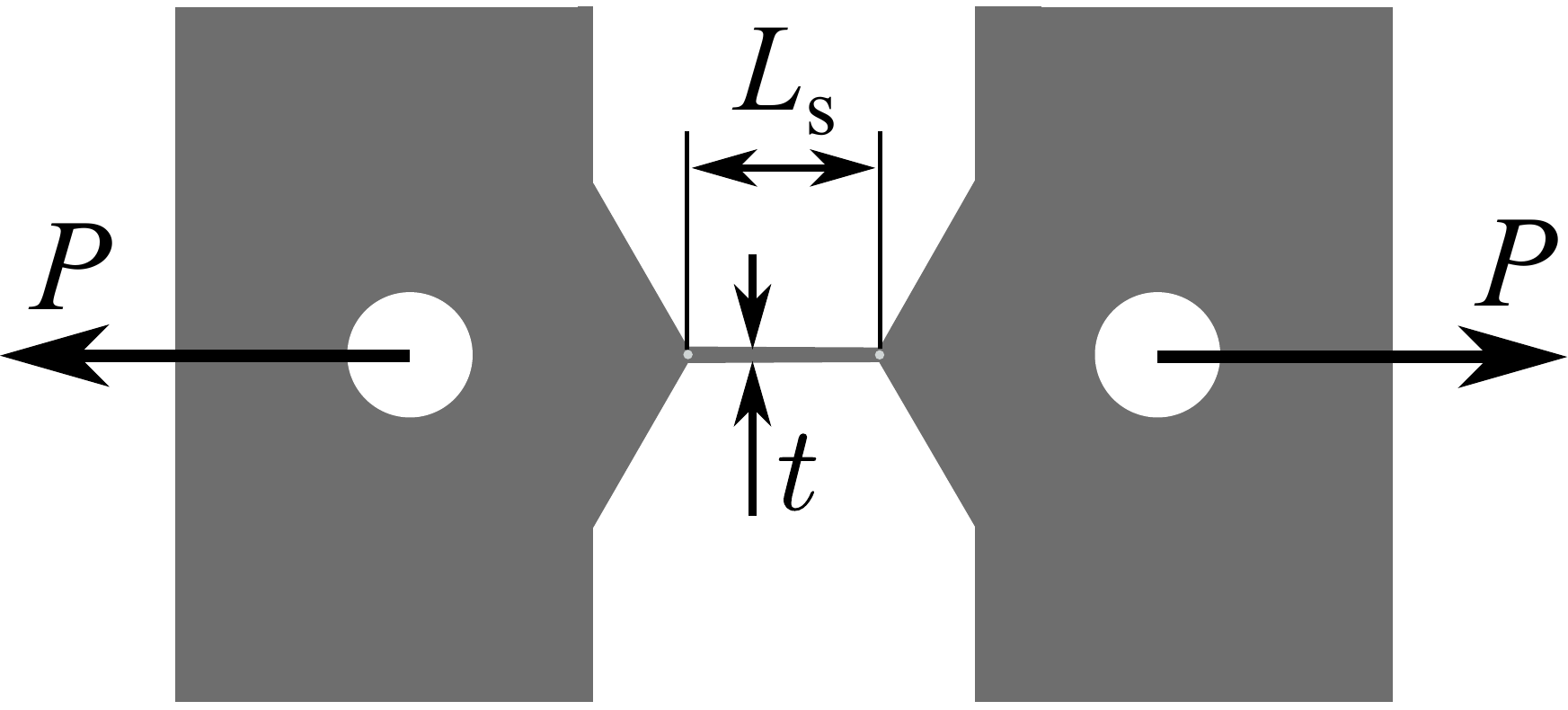}\label{fig:sketch_single_strut}}\\[\baselineskip]
  \sidesubfloat[]{\includegraphics[width=0.47\textwidth]{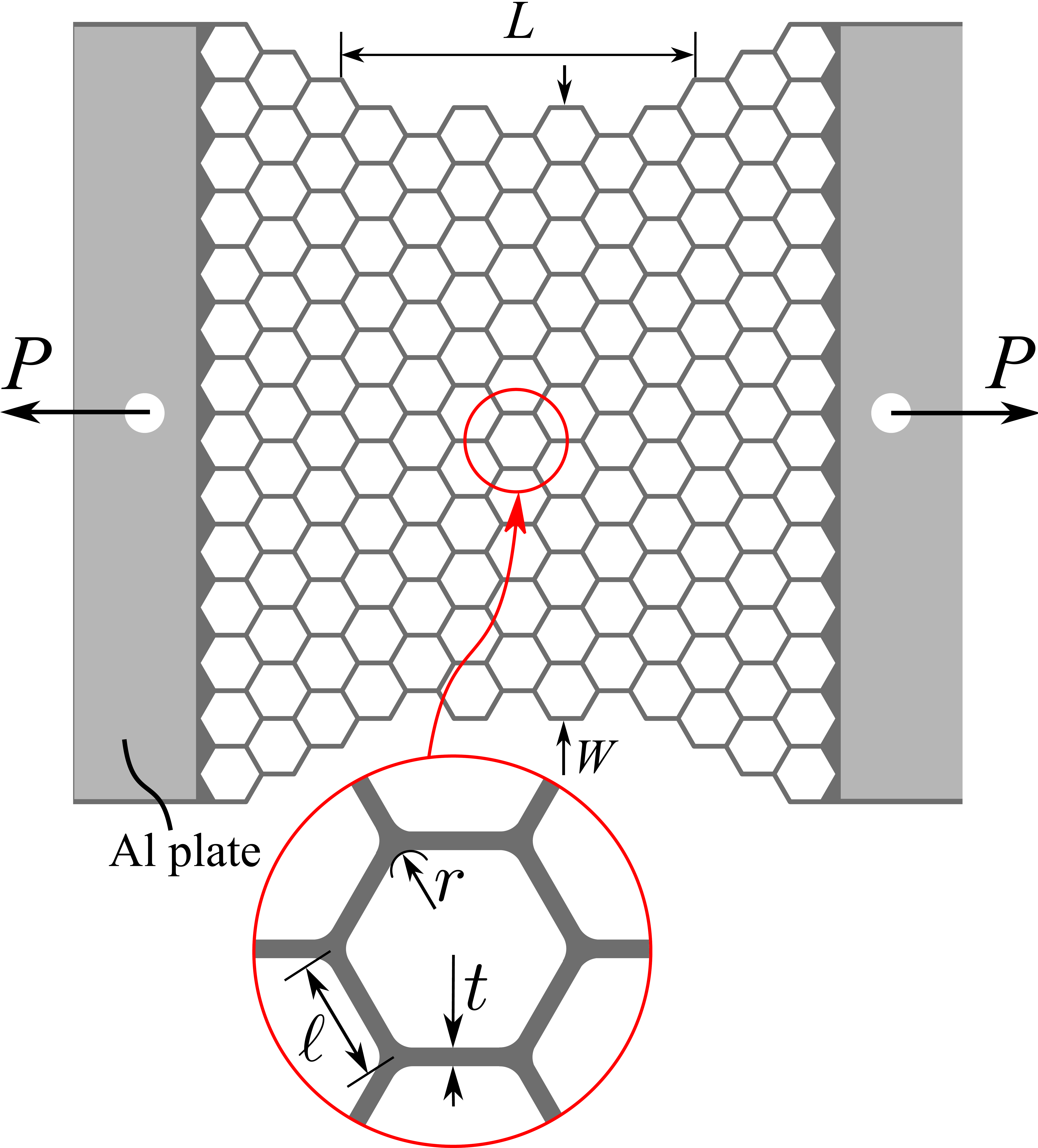}\label{fig:PerfectLatticeSketch}}
  \caption{\subref{fig:sketch_single_strut} Single strut specimen, and
    \subref{fig:PerfectLatticeSketch} lattice specimen of
    $\bar{\rho}=0.11$. The sheet thickness of all samples is
    $B_0=\SI{5}{mm}.$}
  \label{fig:specimens}
\end{figure}

\begin{figure}[htbp]
  \centering \setlength{\labelsep}{-0.25cm}
  \sidesubfloat[]{\includegraphics[width=0.47\textwidth]{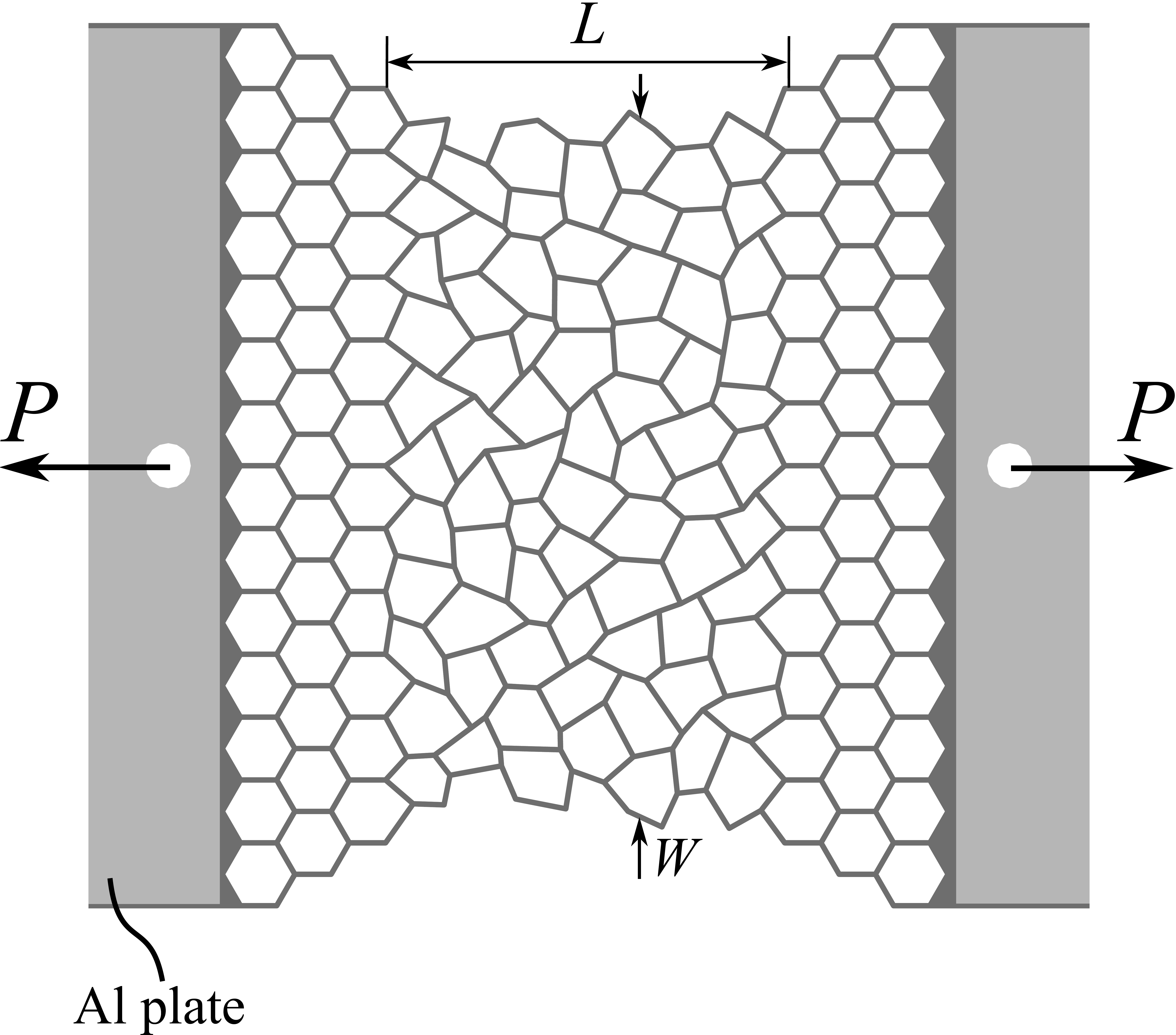}\label{fig:hexLattice_2}}
  \sidesubfloat[]{\includegraphics[width=0.47\textwidth]{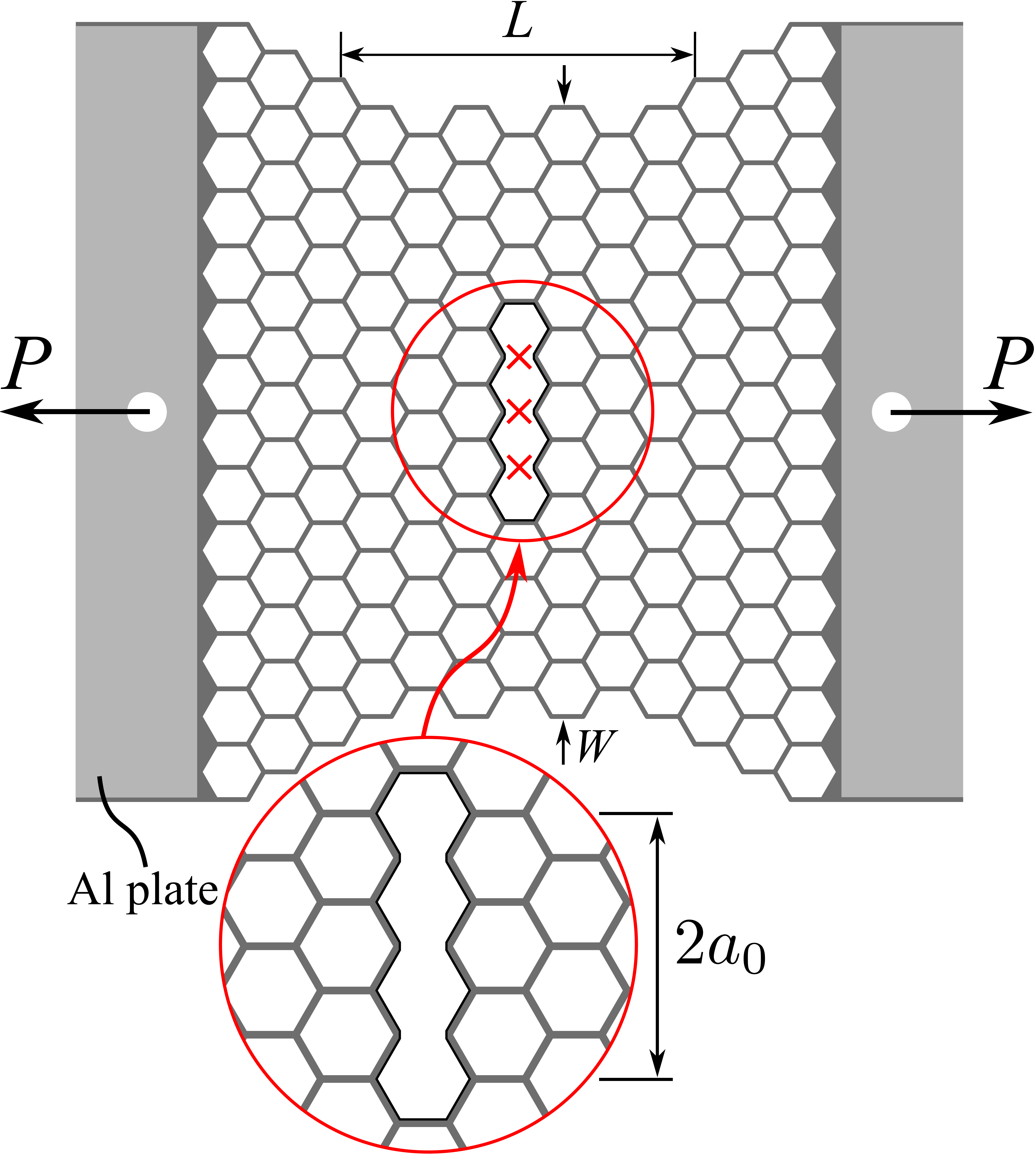}\label{fig:hexLattice_3}}\\[\baselineskip]
  \sidesubfloat[]{\includegraphics[width=0.47\textwidth]{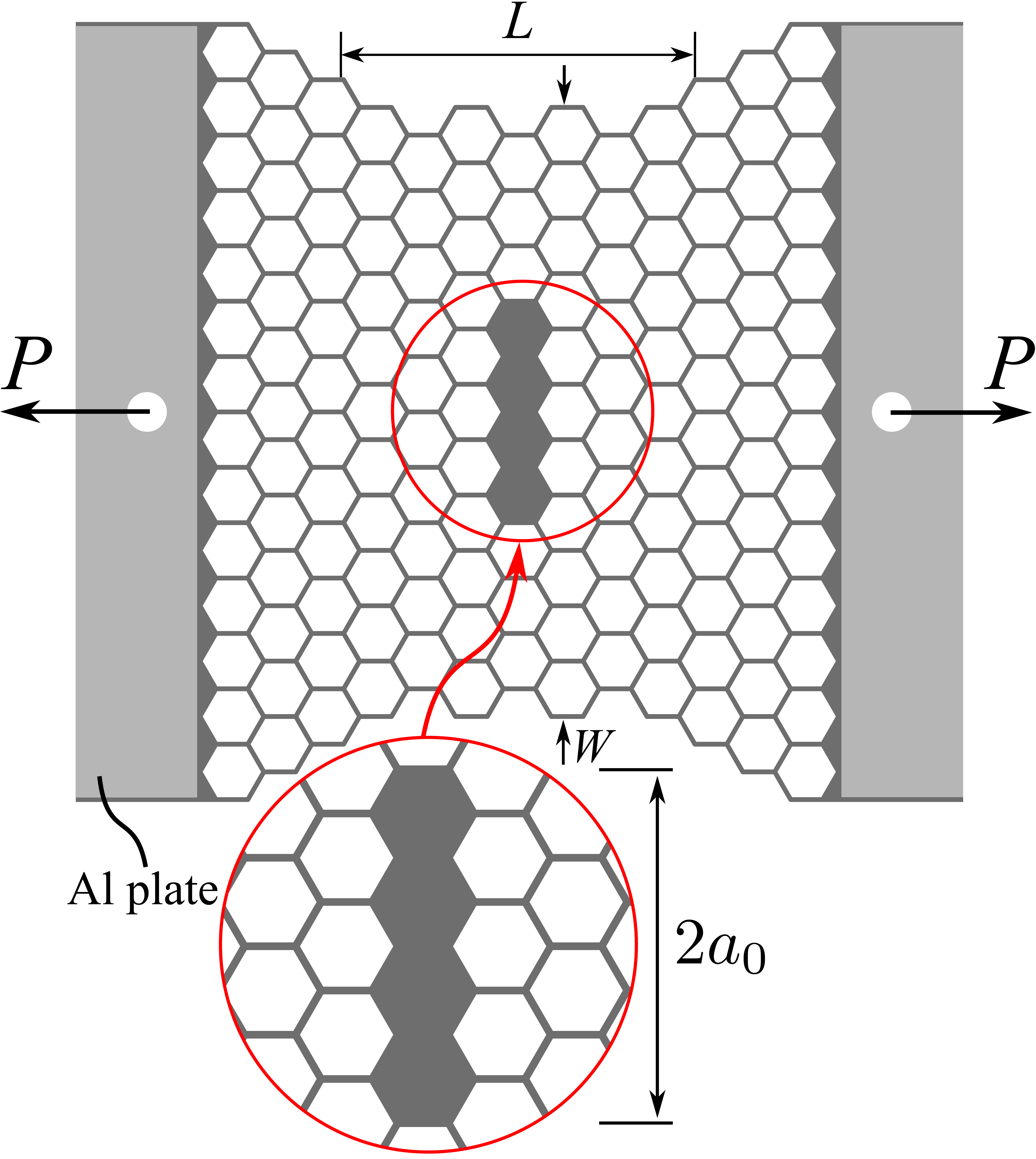}\label{fig:hexLattice_4}}
  \caption{Lattice specimens ($\bar{\rho}=0.11$) containing
    as-designed defects in the form of \subref{fig:hexLattice_2}
    randomly perturbed joints ($R/\ell=0.5$),
    \subref{fig:hexLattice_3} a row of missing cell walls, and
    \subref{fig:hexLattice_4} a row of solid inclusions.}
  \label{fig:irrHexLattice}
\end{figure}

The experimental procedure, including the manufacturing parameters for
laser-cutting the PMMA sheets, has already been detailed in the
companion study~\cite{Seiler2019a}. The same manufacturing routine was
applied in this study, along with the same material batch of cast
\SI{5}{mm} thick PMMA sheets. In brief, specimens were manufactured by
laser-cutting\footnote{\textit{HPC Laser Ltd LS6090 Pro} 80 Watt laser
  cutter; process parameters: \SI{8}{mm/s} cutting speed, 60\% power,
  55\% corner power.} into the following 5 geometries:
\begin{itemize}
\item[(i)] single strut specimen, as shown in
  \cref{fig:sketch_single_strut}, for material characterisation on a
  small scale;
\item[(ii)] regular hexagonal lattice, as shown in
  \cref{fig:PerfectLatticeSketch}, to measure the lattice response,
  absent as-designed defects;
\item[(iii)] irregular hexagonal lattice containing as-designed
  defects in the form of (a) randomly perturbed joints, (b) missing
  cell walls, or (c) solid inclusions, see \cref{fig:irrHexLattice}.
\end{itemize}

The dispersion in strut thickness $t$ and in Plateau border radius $r$
for lattice specimens of types (ii) and (iii) have already been
characterised by X-ray computed tomography (CT) in
Ref.~\cite{Seiler2019a}. Both $t$ and $r$ are defined in
\cref{fig:PerfectLatticeSketch}. The strut thickness was measured at
mid-length of 453 struts. It follows a normal distribution with a mean
value $\bar{t}=\SI{0.47}{mm}$, where the overbar throughout this study
denotes the average value (with the exception of density for which an
overbar denotes relative density). The standard deviation of the strut
thickness is $t_\mathrm{sd}=\SI{0.09}{mm}$. The Plateau border radius
was quantified by a measured mean value $\bar{r}=\SI{0.4}{mm}$ and
standard deviation $r_\mathrm{sd}=\SI{0.1}{mm}$.

For lattice specimens of type~(ii), the relative density of the
lattice $\bar\rho$ is varied from 0.07 to 0.19 by varying the strut
length $\ell$ from \SI{3.0}{mm} to \SI{7.5}{mm}. The PMMA employed in
this study has a glass transition temperature\footnote{The value of
  $T_\mathrm{g}$ was measured by Dynamic Mechanical Analysis (DMA) of
  a single PMMA cantilever beam at an excitation frequency equal to
  \SI{0.1}{Hz} and a heating rate of \SI{5}{\celsius/min}, refer
  to~\cite{VanLoock2018} for details of the test procedure.}
$T_\mathrm{g}= \SI{385}{K}$. All specimens were tested at
$T=\sic{100}=0.97\,T_\mathrm{g}$.

\section{Manufacture of lattice specimens and test method}
\label{sec:manufacture_ref_hex}
A computer-aided drawing (CAD) of the geometry of a regular hexagonal
lattice, as shown in \cref{fig:PerfectLatticeSketch}, was created
using the
\textit{OpenSCAD}\footnote{\textit{https://www.openscad.org}}
software. This CAD file provides an input to the laser cutting machine
with sufficient data to define the translation of the cutting head
relative to a fixed position on the PMMA sheet. The hexagonal lattices
were manufactured to dogbone shape in order to ensure that failure
occurs within the gauge section, see
\cref{fig:PerfectLatticeSketch}. All lattice specimens have a gauge
width $W=11\sqrt{3}\ell$ (or 11 cells) and a gauge length $L=11\ell$
(or 7 cells). The front and back faces of the end, gripping portions
of the lattice specimens were adhered to \SI{2}{mm} thick aluminium
alloy end tabs, see \cref{fig:PerfectLatticeSketch}.

All lattice specimens were tested in uniaxial tension using a servo
hydraulic test machine at a nominal strain rate of
$\dot\varepsilon=\SI{4E-4}{s^{-1}}$. The tests were conducted at
\sic{100} in a temperature-controlled chamber with a glass window for
in-situ observation of the specimen during the test. The temperature
was maintained to a precision of $\Delta T \pm \SI{2}{K}$ during each
test. The load $P$ was measured via a load cell clamped to the
stationary platen of the rig while the extension $u$ of the gauge
length was determined by Digital Image Correlation (DIC), as described
in~\cite{Seiler2019a}. Prior to the start of each test, the lattice
specimens were coated with a thin layer of white chalk, and a speckle
pattern was generated by the spraying of black paint in order to
enhance the contrast of the DIC imagery.  A digital
camera\footnote{maximum resolution: 4608\,$\times$\,3288 pixels,
  \SI{55}{mm} lens for lattices and \SI{100}{mm} lens for single strut
  tests} was used to track facets of size $20\times 20$ pixels in the
vicinity of all nodes to enable sub-pixel resolution of the nodal
displacement.

\section{Material characterisation}
\label{sec:solid_material_response}
The as-manufactured material properties of solid PMMA were measured
from the tensile response of laser-cut single strut specimens of mean
strut thickness $\bar{t}=\SI{0.47}{mm}$ and strut length
$L_\mathrm{s}=\SI{10}{mm}$. The tests were conducted at \sic{100} in a
temperature-controlled chamber as described before. The extension of
the single strut specimens was measured by optical tracking of white
dots at the ends of the gauge length along the centre-line of the
specimen using DIC technique.

The measured nominal stress versus nominal strain response for 3
different strain rates within the range $\SI{4E-4}{s^{-1}}$ to
$\SI{4E-3}{s^{-1}}$ are shown in
\cref{fig:StressStrain_Strut_Images_100C}. The flow strength increases
with increasing strain rate. Negligible necking was found at \sic{100}
due to the visco-plastic material behaviour which is in agreement with
the observations of \citet{VanLoock2018}. The single strut has a
nominal failure strain $\varepsilon_\mathrm{fs}$ of 2.5, or
equivalently, a true failure strain of 1.3. The single strut failure strain is insensitive to
strain rate over the range tested.

\section{Measured tensile response of as-manufactured lattice
  specimens}
\label{sec:lattice_specimens}
\begin{figure}[htbp]
  \centering \setlength{\labelsep}{-1cm}
  \sidesubfloat[]{\includegraphics[width=0.53\textwidth]{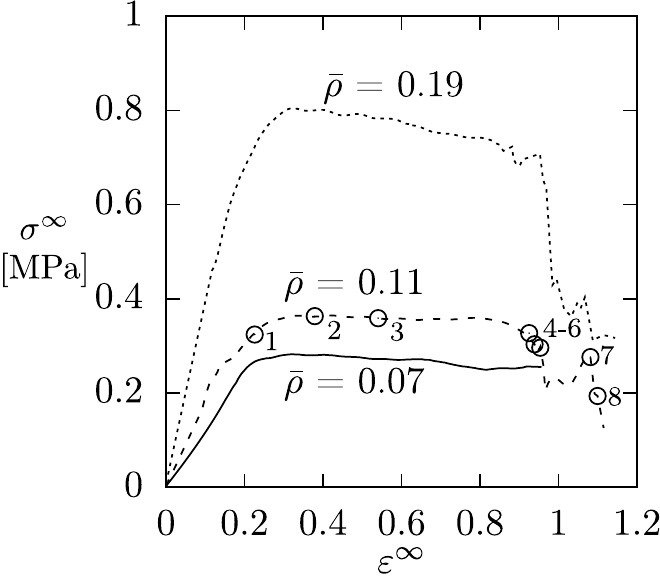}\label{fig:StressStrain_HexLatticePerfect_100C_smoothedWithFailure2}}\hfill
  \setlength{\labelsep}{0.2cm}
  \sidesubfloat[]{\includegraphics[width=0.4\textwidth]{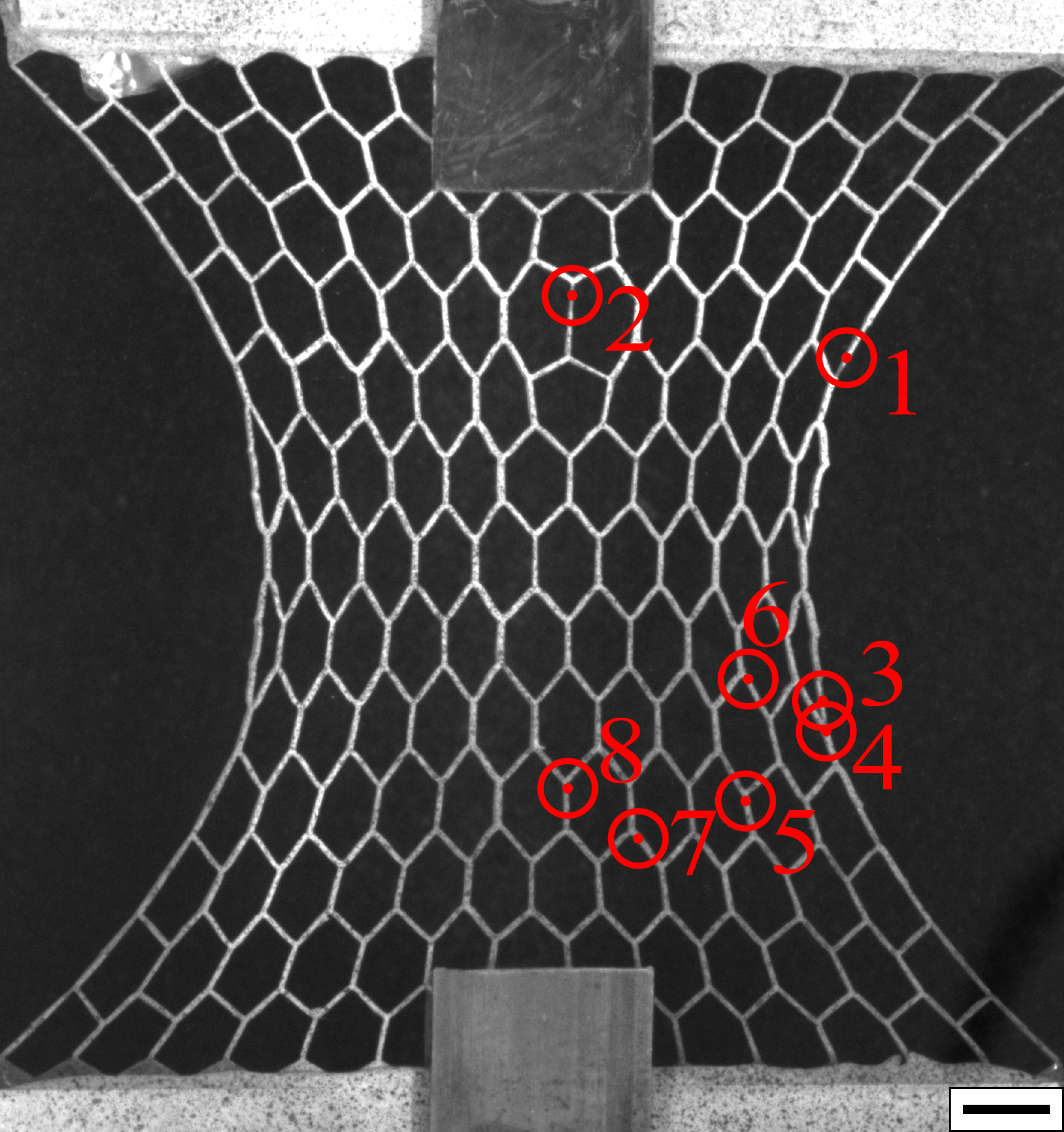}\label{fig:Sample_0039}}
  \caption{\subref{fig:StressStrain_HexLatticePerfect_100C_smoothedWithFailure2}
    Measured macroscopic nominal stress versus nominal strain response
    of lattices for selected values of relative density $\bar\rho$,
    \subref{fig:Sample_0039} deformed lattice ($\bar{\rho}=0.11$)
    before first strut failure along with the sequence and location of subsequent strut failure. The scale bar is of length \SI{10}{mm}.}
  \label{fig:experiment_reg_lattice}
\end{figure}

The measured macroscopic nominal stress $\sigma^\infty$ versus nominal
strain $\varepsilon^\infty$ responses of 3 lattice specimens of
\cref{fig:PerfectLatticeSketch} are shown in
\cref{fig:StressStrain_HexLatticePerfect_100C_smoothedWithFailure2}. Each
specimen contains 453 struts of length
$\ell=\SI{5}{mm}$. Representative curves are shown for a single
specimen at each relative density of $\bar{\rho}=0.07$, $0.11$, and
$0.19$. The macroscopic nominal quantities $\sigma^{\infty}$ and
$\varepsilon^\infty$ are defined in terms of the measured load $P$ and
the extension $u$ of the gauge length as $\sigma^{\infty}=P/(WB_0)$
and $\varepsilon^\infty=u/L$, respectively. The dimensions $(W,L)$ are
defined in \cref{fig:PerfectLatticeSketch}; $W$ equals \SI{96}{mm},
$L$ equals \SI{55}{mm}, and the out-of-plane thickness $B_0$ equals
\SI{5}{mm}. 

The sequence of strut failure in a lattice specimen of relative
density $\bar{\rho}=0.11$ is shown in
\cref{fig:StressStrain_HexLatticePerfect_100C_smoothedWithFailure2},
with the corresponding locations of strut failure marked in
\cref{fig:Sample_0039}. The following observations are made from
\cref{fig:StressStrain_HexLatticePerfect_100C_smoothedWithFailure2,fig:Sample_0039},
and are representative of the response for all 3 values of
$\bar{\rho}$. All failed struts exist within the gauge section of the
specimen and strut failure occurs always close to a joint. First strut
failure does not necessarily occur at the edge of the
specimen. Subsequent strut failures occur at random locations, leading
to a diffuse mode of damage, see \cref{fig:Sample_0039}. First strut
failure is accompanied by only a small drop in the load $P$ (and in
turn $\sigma^\infty$). The cell wall solid has a sufficiently high
strain-rate hardening response that the macroscopic stress
$\sigma^\infty$ is almost constant during the progressive failure of
the first 6 struts at random locations within the
lattice. Approximately 50\% of the failed struts are inclined at
$\pm 60^\circ$ to the loading axis in the undeformed configuration. In
contrast, 90\% of the failed struts of elastic-brittle PMMA lattices
were inclined at $\pm 60^\circ$ to the loading
axis~\cite{Seiler2019a}. \changes{It is appreciated that this specimen is not sufficiently long
  in order to give the response of an infinite domain of lattice
  material. The large Poisson contraction and large tensile strain ($\varepsilon^\infty \approx 1$)
  exhibited by the specimen lead to a non-uniform stress state. This
  is difficult to overcome in an experimental manner: a specimen of
  gauge width on the order of 100 cells (or \SI{1}{m}) would be needed
  to give a uniform stress state. Thus, an alternative strategy of
  modelling was employed: the full specimen was analysed by finite element simulations
  in order to make predictions, instead of the  periodic cell calculations.}

\begin{figure}[htbp]
  \centering
  \includegraphics[width=0.5\textwidth]{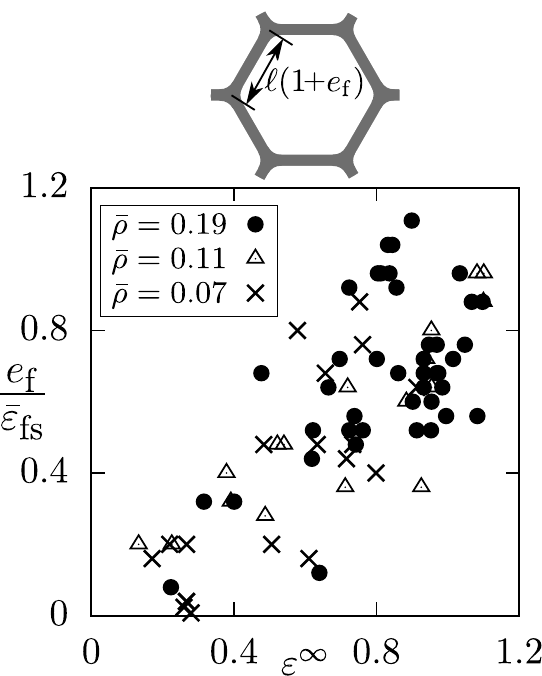}
  \caption{Measured nominal axial failure strain of struts
    $e_\mathrm{f}$ versus the macroscopic nominal strain
    $\varepsilon^\infty$.}
  \label{fig:LocalStrainFailure_RemoteStrainFailure_perfectHex_ComparedHarika2}
\end{figure}

It is instructive to compare the strut nominal ductility
$e_\mathrm{f}$ with the nominal failure strain of the cell wall solid
$\bar{\varepsilon}_\mathrm{fs}$, as follows. The nominal axial strain
$e$ of each strut was measured in the gauge section of the specimens,
and the axial failure strain $e_\mathrm{f}$ of each strut was recorded
just prior to the strut failure. This failure strain $e_\mathrm{f}$ is
the nominal value over the length of a strut, as measured from the
centre point of the joints at each end of the failed strut. This
definition is shown in the inset of
\cref{fig:LocalStrainFailure_RemoteStrainFailure_perfectHex_ComparedHarika2}.
A wide dispersion in strut ductility $e_\mathrm{f}$ was observed
\changes{in both the inclined and aligned struts with regard to the
  loading direction.} To illustrate this, the measured value of
$e_\mathrm{f}$ was plotted as a function of macroscopic nominal strain
$\varepsilon^\infty$ in
\cref{fig:LocalStrainFailure_RemoteStrainFailure_perfectHex_ComparedHarika2}. The
data from 3 nominally identical specimens are shown, for each value of
$\bar{\rho}$. The strut ductility $\eavg$ approaches the value of
$\bar{\varepsilon}_\mathrm{fs}=2.5$ at a macroscopic nominal strain
$\varepsilon^\infty$ exceeding 0.8.

\section{Simulations}
\label{sec:numerical_approach}

\subsection{Details of the FE model}
\begin{figure}[htbp]
  \centering \setlength{\labelsep}{-0.5cm}
  \sidesubfloat[]{\includegraphics[width=0.99\textwidth]{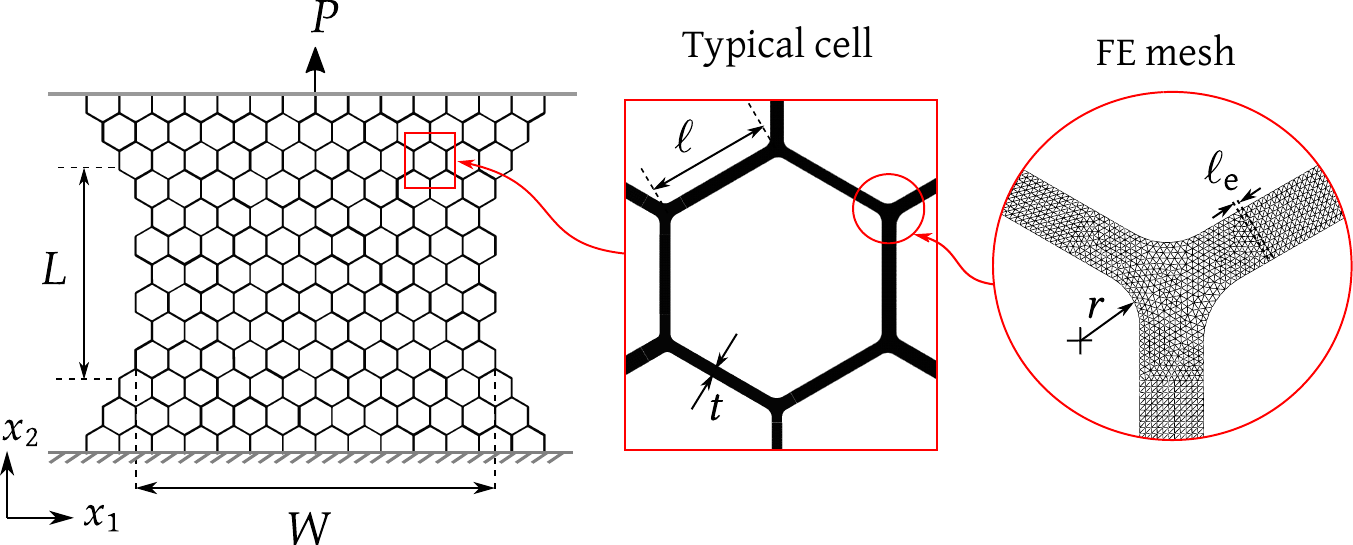}\label{fig:fe_mat_sketch_lattice_v2a}}\\[\baselineskip]
  \setlength{\labelsep}{0cm}
  \sidesubfloat[]{\includegraphics[width=0.6\textwidth]{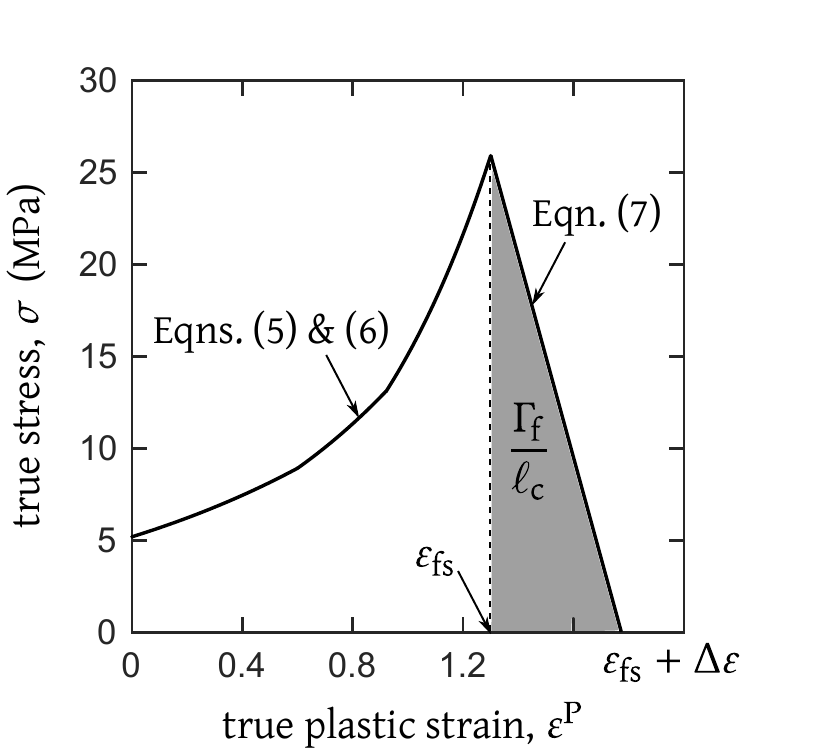}\label{fig:fe_mat_sketch_lattice_v2b}}
  \caption{Details of the FE model:
    \subref{fig:fe_mat_sketch_lattice_v2a} geometry of an as-manufactured lattice specimen along with the loading and
    boundary conditions employed in the FE simulations. A typical cell within the
    lattice is shown along with the FE mesh at a
    joint. \subref{fig:fe_mat_sketch_lattice_v2b} Assumed true stress
    versus true plastic strain response of the cell wall solid for $\dot\varepsilon=\SI{4e-4}{s^{-1}}$.}
 \label{fig:fe_sketch}
\end{figure}

The finite element (FE) method is used to investigate numerically the
tensile creep response of PMMA lattices made from rapid
prototyping. The main objective of the FE study is to examine the
sensitivity of the macroscopic response to as-manufactured defects
such as the dispersion in strut thickness, Plateau border radius, and
strut ductility. Two geometries of lattice specimen of relative
density $\bar{\rho}=0.11$ are modelled explicitly for comparison with
the experimental observations: (i) an ideal lattice with constant
strut thickness and constant Plateau border radius, and (ii)
structural realisations of the as-manufactured lattice, as shown in
\cref{fig:fe_sketch}, for direct comparison with the experimental
observations. The specimen geometry was scanned by CT and the FE mesh
was constructed using the centre plane of the scanned geometry. The
struts within the lattice have a variable thickness, with a mean value
of $\bar{t}=\SI{0.47}{mm}$ and a standard deviation of
$t_{\rm \rms}/\bar{t}=0.19$. Quasi-static calculations were performed
with \textsc{Abaqus}/Explicit v6.14 to simulate the deformation and
failure response of these specimens under remote uniaxial tensile
loading.

The FE mesh comprises quadratic triangular elements in plane strain
(type CPE6M) of uniform size $\ell_{\rm e}$ such that the thinnest
strut in the lattice has at least five elements across its thickness,
and the stress concentration at the Plateau borders is adequately
captured, as shown in the inset of
\cref{fig:fe_mat_sketch_lattice_v2a}. Uniaxial loading of the specimen
is simulated by constraining all degrees of freedom along the bottom
edge of the specimen $(u_1=u_2=0)$ while the top edge is displaced in
the $x_{2}$-direction of the specimen $(u_1=0, u_2=u)$, see
\cref{fig:fe_mat_sketch_lattice_v2a}.

\subsubsection{Material model}
The cell wall material is modelled as an elastic, visco-plastic
solid. Tensile fracture of the individual struts of the lattice is
simulated using an idealised Johnson-Cook type continuum damage
mechanics approach. This approach assumes that damage initiates in
accordance with a local strain criterion whereas damage evolves on the
basis of a prescribed work of fracture. A detailed explanation of the
material model assumed in the FE simulations is given below.

\begin{figure}[htbp]
  \centering
  \includegraphics[width=0.5\textwidth]{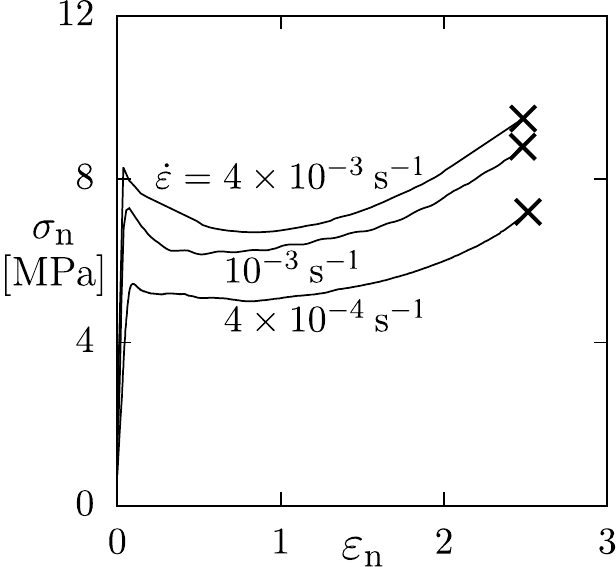}
  \caption{Nominal stress versus nominal strain response of single
    strut samples at \sic{100}, for 3 selected values of strain rate.}
  \label{fig:StressStrain_Strut_Images_100C}
\end{figure}

The cell wall has a Young's modulus $E_{\rm s}=\SI{400}{MPa}$ and
Poisson's ratio $\nu_{\rm s}=0.3$, as measured form the single strut
specimens. We adopt a visco-plastic constitutive law in terms of the
true plastic strain $\ep$, true stress $\sigma$, and strain rate
$\eprate$. It has the form
\begin{equation}
  \sigma = \sigma_0 \; \textrm{sinh}^{-1} \left( \dfrac{\dot\varepsilon}{\dot{\varepsilon}_{\rm 0}} \right) \; f\left( \varepsilon^{\rm P}   \right)
\label{eqn:viscolaw}
\end{equation}
where the reference values $(\sigma_0, \dot{\varepsilon}_{\rm 0})$ are
taken as $\sigma_0 = \SI{1}{MPa}$ and
$\dot{\varepsilon}_{\rm 0}= \SI{4.4e-6}{s^{-1}}$ such that
$\sigma \left( \dot{\varepsilon}, \varepsilon^{\rm P}=0 \right) =
\SI{5.2}{MPa}, \SI{6.1}{MPa}$ and $\SI{7.5}{MPa}$ for the 3 values of
strain rate $\erate=\SI{4e-4}{s^{-1}}$, \SI{e-3}{s^{-1}} and
\SI{4e-3}{s^{-1}}, respectively. It remains to specify
$f\left( \varepsilon^{\rm P} \right)$ in \eqref{eqn:viscolaw}. A
curve-fitting procedure is carried out on the measured $\sigma$ versus
$\ep$ data excluding the initial peak and the subsequent softening
characteristics of \cref{fig:StressStrain_Strut_Images_100C}. This
procedure results in the following functional form for
$f\left( \varepsilon^{\rm P} \right)$:

\begin{equation}
  f\left( \varepsilon^{\rm P}   \right)=  
  \begin{cases} \exp(0.9\varepsilon^{\rm P}) & \quad \mbox{for } 0 \leq \varepsilon^{\rm P} < 0.6\\               
    0.835 \exp(1.2\varepsilon^{\rm P}) & \quad  \mbox{for } 0.6 \leq \varepsilon^{\rm P} < 0.92\\
    0.48 \exp(1.8\varepsilon^{\rm P}) & \quad  \mbox{for } \varepsilon^{\rm P} \geq 0.92 \end{cases} 
\label{eqn:stressstrain_fe}
\end{equation}

Following \Citet{Cook1985}, it is assumed that damage at a material
element (i.e. integration point) \textit{initiates} when the
equivalent plastic tensile strain at that point attains the solid
tensile ductility, $\varepsilon_{\rm fs}$, as indicated in
\cref{fig:fe_sketch}(b). A deterministic value of strut ductility
(i.e. true strain to failure), $\varepsilon_{\rm fs}=1.3$, is assumed
for all struts in the lattice taken from the measured mean ductility
of single strut specimens, recall
\cref{fig:StressStrain_Strut_Images_100C}.

The subsequent \textit{evolution} of damage at a material point is
specified via a linear softening $\sigma$ versus $\ep$ relationship
given by
\begin{equation}
  \sigma = \sigma_{\rm f} \left( 1 - \dfrac{\varepsilon^{\rm P}-\varepsilon_{\rm fs}}{\Delta\varepsilon}   \right)
\label{eqn:damagelaw}
\end{equation}
where $\sigma_{\rm f}$ is the stress level at
$\varepsilon^{\rm P}=\varepsilon_{\rm fs}$ as given by
\eqref{eqn:stressstrain_fe}, and $\Delta\varepsilon$ is the plastic
strain increment over the softening portion of the response, see
\cref{fig:fe_sketch}(b). The value of $\Delta\varepsilon$ follows from
the specified work of fracture in the softening regime,
$\Gamma_{\rm f}$, and the characteristic length associated with the
finite element, $\ell_{\rm c}$, as:
$\Delta\varepsilon = 2\Gamma_{\rm f}/\sigma_{\rm f}\ell_{\rm c}$. Note
that $\ell_{\rm c} = \ell_{\rm e}/2$ for quadratic triangular finite
elements. A value of $\Gamma_{\rm f}= 2.5 \ \rm kJ/m^{2}$ is assumed
in the current FE simulations upon making use of the Irwin relation
$\Gamma_{\rm f}= K_{\rm IC}^2/E_\mathrm{s}$ with
$K_{\rm IC}=\SI{1}{MPa\sqrt{m}}$ as taken from \cite{Choi1993} and
$E_\mathrm{s}=\SI{400}{MPa}$ from the single strut experiments. It is
noted that the specification of damage evolution via
\eqref{eqn:damagelaw} in terms of the size of the finite element
alleviates the problem of mesh size dependence of the solution; refer
to Ref.~\cite{Oliver1989} for details. \changes{We emphasise that this
  Johnson-Cook type failure model gives rise to strut failure in a combined
  stretching and bending manner at a location near the joints. This is
  in agreement with the observed nature of strut failure.}

\subsection{FE predictions of the tensile response of as-manufactured
  lattice specimens}
\begin{figure}[ht]
  \centering \setlength{\labelsep}{-0.5cm}
  \sidesubfloat[]{\includegraphics[width=0.44\textwidth]{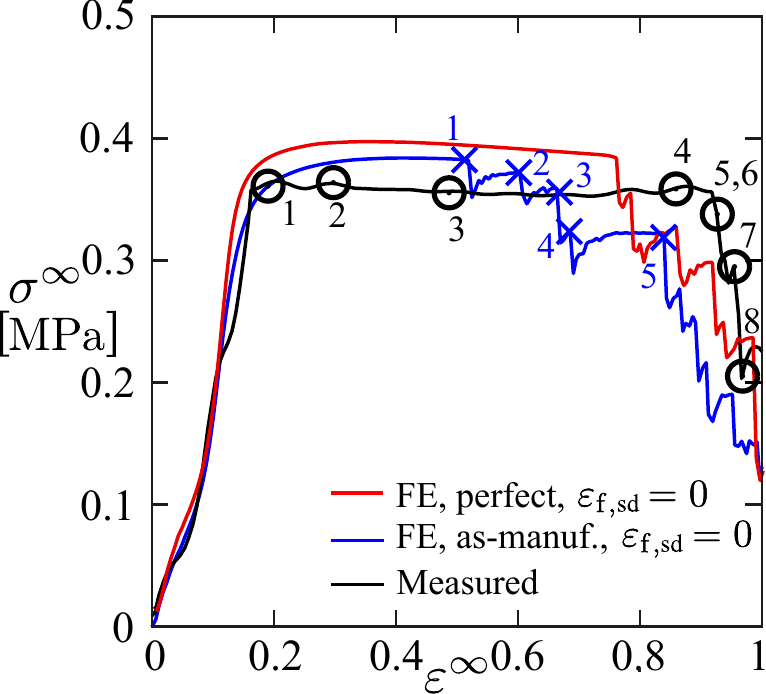}\label{fig:macroscopicresponse}}\hfill
  \addtocounter{subfigure}{1}
  \sidesubfloat[]{\includegraphics[width=0.44\textwidth]{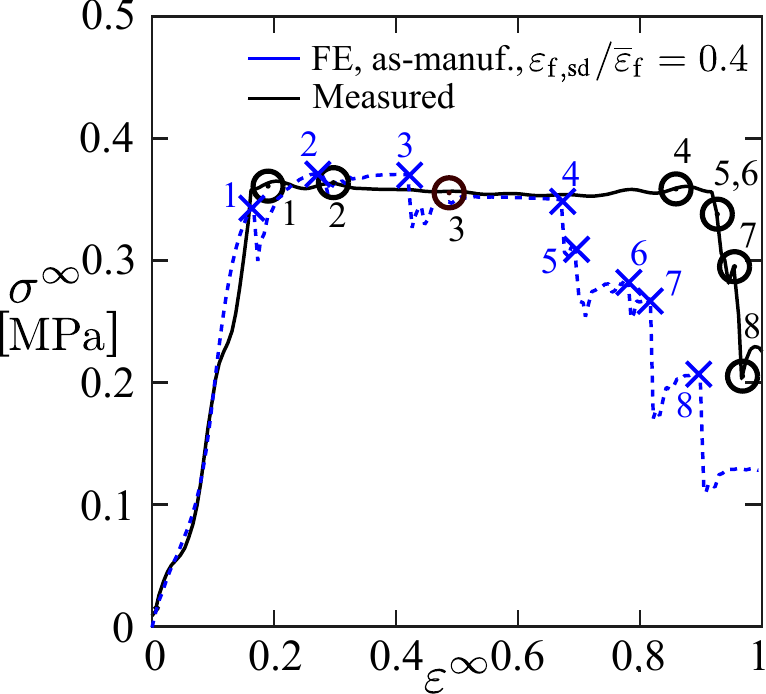}\label{fig:macroscopicresponse_2}}\\[\baselineskip]
  \setlength{\labelsep}{0.2cm} \addtocounter{subfigure}{-2}
  \sidesubfloat[]{\includegraphics[width=0.95\textwidth]{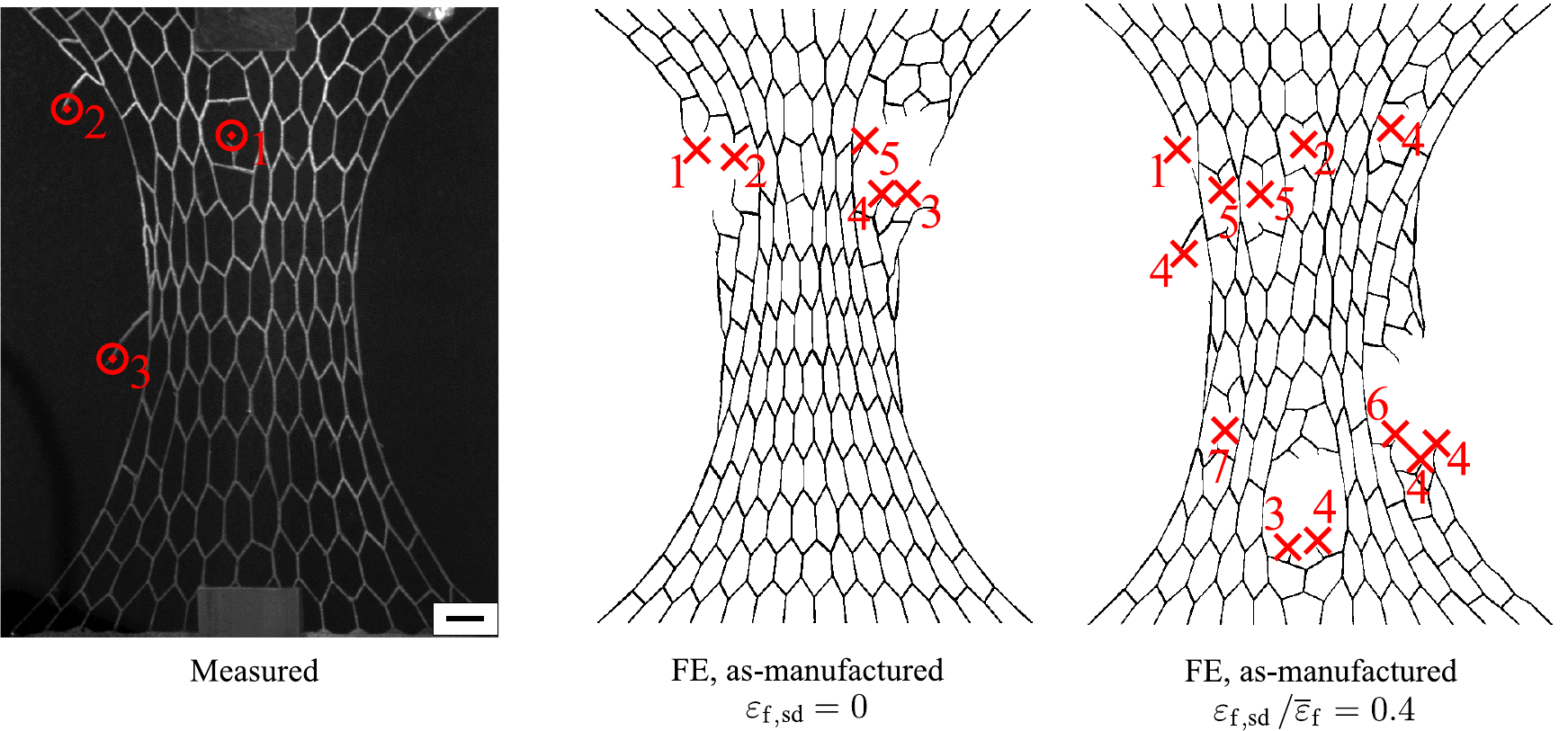}\label{fig:damage}}
  \caption{Measured versus predicted response for a lattice of
    relative density $\bar{\rho}=0.11$:
    \subref{fig:macroscopicresponse} macroscopic nominal stress
    $\sinf$ versus macroscopic nominal strain of perfect lattices (FE)
    and one realisation of the as-manufactured lattice (FE and measured);
    \subref{fig:damage} damage state at
    $\varepsilon^\infty=0.8$. Circles denote the location of strut failure in experiments while crosses denote the predicted failure sites. The numbers in each case denote the position of the strut in the failure sequence. The scale bar is of length \SI{10}{mm}.
    \subref{fig:macroscopicresponse_2} Predicted $\sinf$ versus $\varepsilon^\infty$ response in the presence of a dispersion in strut ductility. Measured response for this specimen is included for comparison.}
\label{fig:fe_response}
\end{figure}

In order to quantify the role of as-manufactured defects, such as the
variation in strut thickness and in Plateau border radius, baseline FE
simulations were performed on a lattice of a uniform strut thickness
$t$, Plateau border radius $r$, and a deterministic value of cell wall
ductility $\ef$. Accordingly, the predicted response of a perfect
lattice specimen of relative density $\bar{\rho}=0.11$ is shown in
\cref{fig:macroscopicresponse} for $t=\bar{t}=\SI{0.47}{mm}$ and
$r=\bar{r}=\SI{0.4}{mm}$ and a true failure strain $\ef=1.3$ for all
struts in the lattice. The measured response of one of the
as-manufactured specimens of $\bar{\rho}=0.11$ is included in
\cref{fig:macroscopicresponse} for comparison. We find from
\cref{fig:macroscopicresponse} that the initial stiffness of the
lattice is predicted accurately by the perfect geometry, implying a
negligible effect of the geometric imperfections on the elastic
stiffness of the lattice. The first strut to fail in the perfect
lattice does so at a macroscopic nominal strain of $0.76$, which significantly
exceeds the measured nominal strain at first strut failure of
$0.18$. The predicted macroscopic strength of the perfect lattice also
exceeds the measured value.

The sensitivity of macroscopic strength to the presence of
as-manufactured defects is explored by performing an FE simulation on
the geometry of the as-manufactured specimen as defined from a CT scan
of the mid-plane section. This geometry contains a dispersion in strut
thickness and in Plateau border radius as specified by
$t_{\rm \rms}/\bar{t}=0.19$ and $r_{\rm \rms}/\bar{r}=0.25$,
respectively. A deterministic value of cell wall ductility $\ef=1.3$
is again assumed for all struts in the lattice such that
$\varepsilon_{\rm fs,\rms}=0$. The predicted response of the
as-manufactured specimen shows early strut failure at a macroscopic
nominal strain of $0.42$ and a macroscopic strength slightly above the
measured value, see \cref{fig:macroscopicresponse}. The failure of
struts in the FE simulations for both the perfect and as-manufactured
geometries occurs in a correlated crack-like manner as shown in
\cref{fig:damage}. This contrasts with the observed diffuse damage
mode. We proceed to show that the diffuse damage mode requires a
dispersion in strut ductility.

\subsubsection{Effect of a dispersion in strut ductility}
\label{sec:effect_spatial_variation_ductility}
\begin{figure}[ht]
  \centering
  \includegraphics[width=0.5\textwidth]{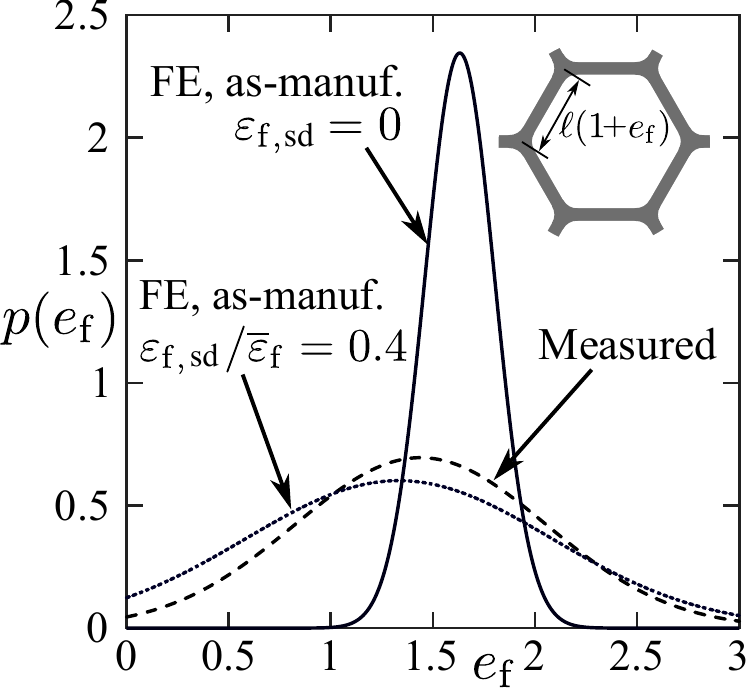}
  \caption{The probability distribution function $p({e}_{\rm f})$ of
    the nominal axial failure strain of struts ${e}_{\rm f}$ for a lattice of
    relative density $\bar{\rho}=0.11$.}
\label{fig:pdfstrutductility}
\end{figure}

The distribution of measured strut nominal ductility $\eavg$ is
re-plotted in 
 \cref{fig:pdfstrutductility} in the form of a probability
distribution curve $p({e}_{\rm f})$, based on the data from 
\cref{fig:LocalStrainFailure_RemoteStrainFailure_perfectHex_ComparedHarika2}. The measured distribution fits a
Gaussian distribution of mean value $\bar{e}_{\rm f}=1.63$ and
standard deviation $e_{\rm \rms}/\bar{e}_{\rm f}=0.4$. Is this
observed distribution in strut ductility due to the imperfect geometry
of the lattice (i.e. dispersion in $t$ and $r$) or is it due to the
scatter in material ductility from one strut to the next? In order to
address this, a set of 5 FE simulations were performed, all based on
the same mesh (as defined by the CT measurement). However, each
simulation assumed a different realisation of randomly-generated strut
ductility $\varepsilon_{\rm fs}$ in accordance with the normal
distribution curve of $\bar{\varepsilon}_{\rm fs}=1.3$ and
$\varepsilon_{\rm fs,\rms}/\bar{\varepsilon}_{\rm f} = 0.4$. Note that
the assumed value of $\bar{\varepsilon}_{\rm fs}$ equals the measured
value of single strut ductility from
\cref{fig:StressStrain_Strut_Images_100C}. The assumed value of
$\varepsilon_{\rm fs,\rms}/\bar{\varepsilon}_{\rm f}$ is taken to
equal the measured value $e_\mathrm{f,sd}/\bar{e}_\mathrm{f}$ of strut
ductility from \cref{fig:pdfstrutductility}.
  
The predicted $\sigma^\infty$ versus $\varepsilon^\infty$ response of
one of the 5 structural realisations is shown in
\cref{fig:macroscopicresponse_2}, along with the measured response of
this specimen. We find from \cref{fig:macroscopicresponse_2} that an
assumed dispersion in strut ductility leads to the early failure of
some struts, consistent with experimental observations. The first
strut fails at the edge of the specimen at a nominal macroscopic
strain of $\einf=0.18$; this strut is of thickness $t=\SI{0.3}{mm}$
and ductility $\ef=1.03$. Subsequent strut failure in the FE
simulation occurs at random locations within the lattice. At
$\varepsilon^\infty=1$, the size of the cluster of failed struts
approaches the half width of the specimen and macroscopic stress drops
to 30\% of the value at first strut failure. The sequence and location
of strut failure for the assumed distribution
${\varepsilon_{\rm fs,\rms}}/\bar{\varepsilon}_{\rm fs} = 0.4$ is
shown in \cref{fig:damage}: the mode of damage is diffuse in nature,
and in good agreement with the experimental observations.

Each FE simulation also gives rise to a sequence of failed struts
under increasing macroscopic nominal strain $\varepsilon^\infty$. The
predicted strut nominal ductility $\eavg$ of each failed strut (from
the 5 simulations) is plotted in the form of a probability
distribution curve in \cref{fig:pdfstrutductility}. It is in excellent
agreement with the observed distribution, implying that the assumed
scatter in material failure strain
${\varepsilon_{\rm fs,\rms}}/\bar{\varepsilon}_{\rm fs} = 0.4$ is of
the correct level. The strong influence of the dispersion in material
ductility from one strut to the next on the creep response of the
lattice becomes evident upon comparing the predicted probability
distribution of $\eavg$ of the same lattice geometry, but with a
deterministic value of strut ductility
($\bar{\varepsilon}_{\rm fs}=\varepsilon_\mathrm{fs}=1.3$ and
${\varepsilon_{\rm fs,\rms}} =0$), as shown in
\cref{fig:pdfstrutductility}: presence of geometric imperfections
alone (i.e. dispersion in $t$ and $r$) in the FE mesh leads to a
narrow distribution in $p(\eavg)$. We conclude that a dispersion in
strut ductility is essential to lead to early strut failure and
diffuse damage prior to catastrophic fracture of the lattice, as
observed in the experiments. In the following section, a discussion is
given of the required material properties that govern the competition
between a diffuse and a correlated mode of strut failure: the primary
properties are the dispersion in strut ductility and the transition
flaw size.

\subsection{Damage tolerance in creep regime}
\begin{figure}[ht]
  \centering
  \sidesubfloat[]{\includegraphics[width=0.42\textwidth]{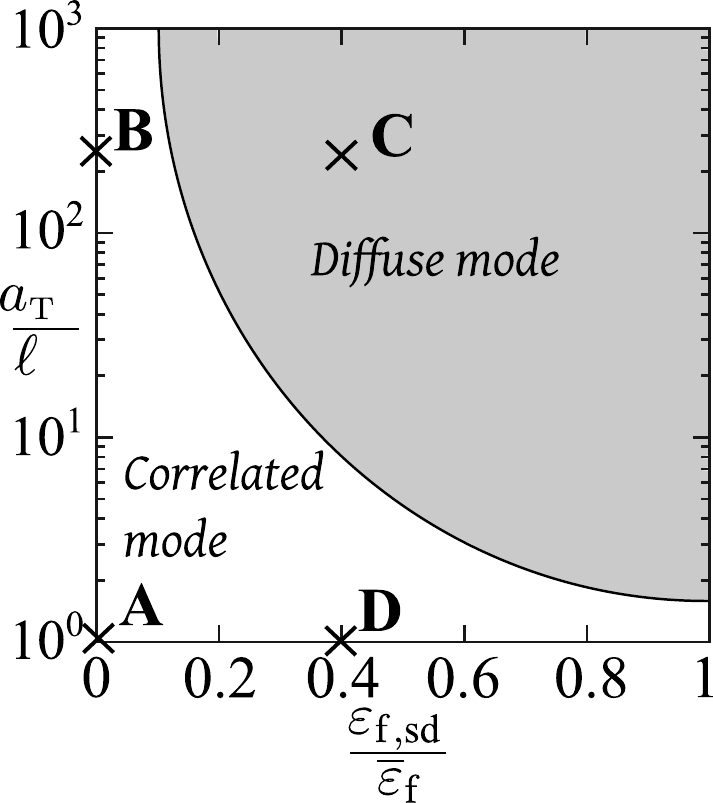}\label{fig:damage_mode}}\hfill
  \sidesubfloat[]{\includegraphics[width=0.45\textwidth]{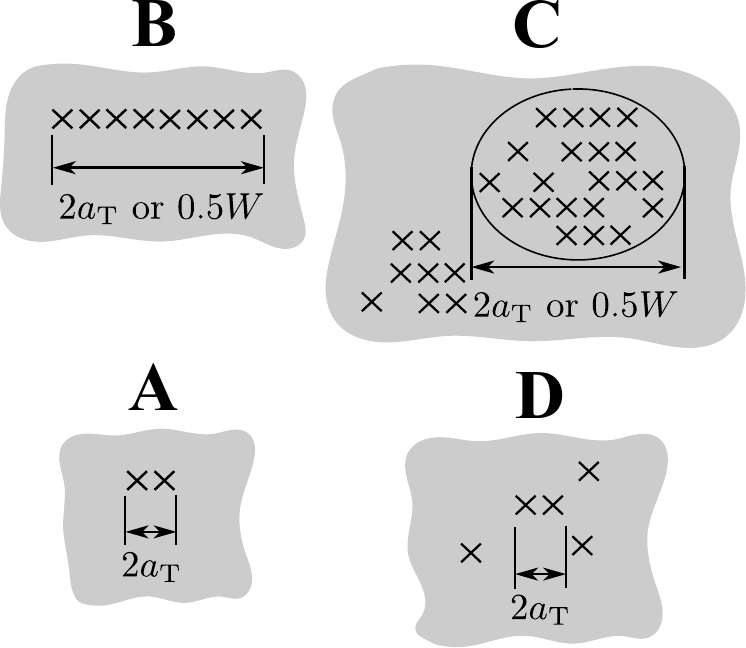}\label{fig:damage_mode_crack}}\\[\baselineskip]
  \setlength{\labelsep}{-0.5cm}
  \sidesubfloat[]{\includegraphics[width=0.47\textwidth]{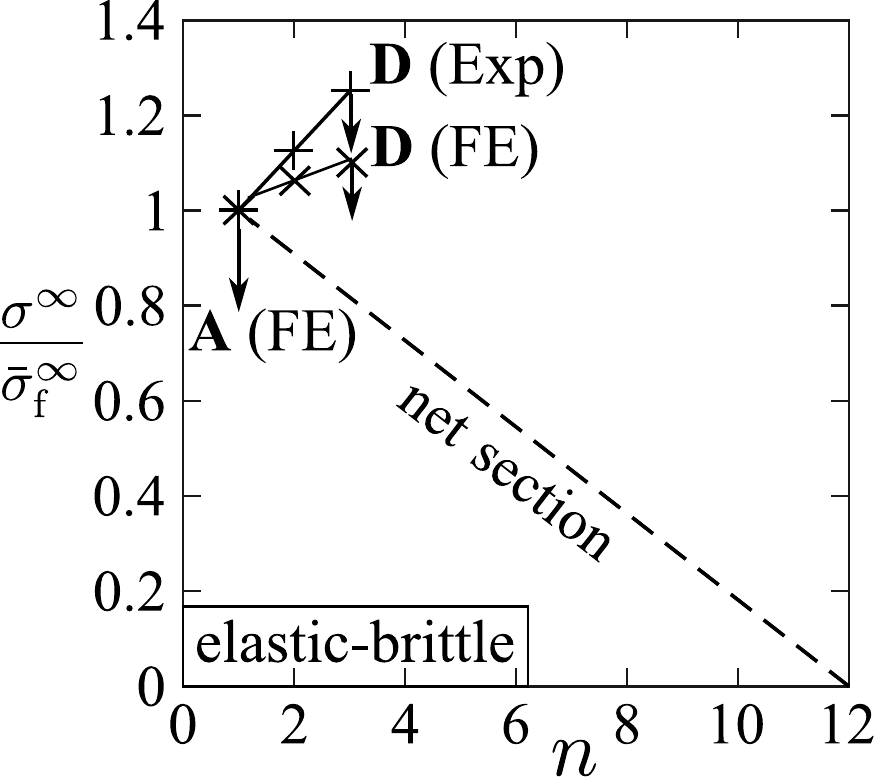}\label{fig:fe_netsection1}}\hfill
  \sidesubfloat[]{\includegraphics[width=0.47\textwidth]{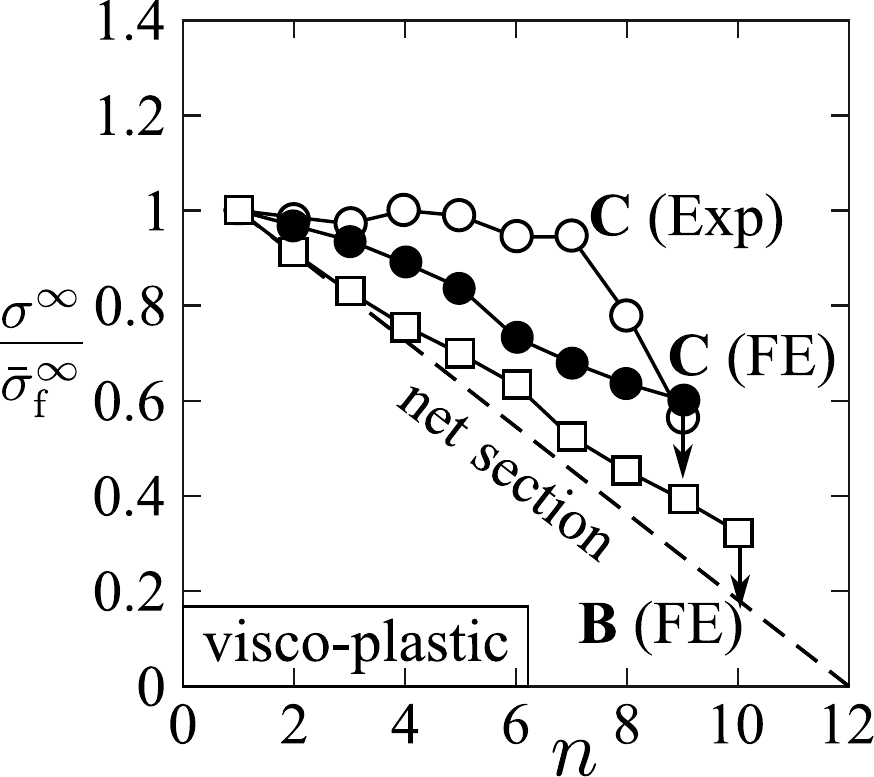}\label{fig:fe_netsection2}}
  \setlength{\labelsep}{0.2cm}
  \caption{Correlated versus diffuse damage: \subref{fig:damage_mode}
    Map showing the parameter space for correlated versus diffuse
    modes of damage as a function of transition flaw size
    $a_{\rm T}/\ell$ and the dispersion in strut ductility
    ${\varepsilon_{\rm f,sd}}/\bar{\varepsilon}_{\rm f}$;
    \subref{fig:damage_mode_crack} damage state at onset of fast
    fracture for cases A through D as labelled in
    \subref{fig:damage_mode}. The predicted and measured macroscopic
    stress during the progressive failure of struts for
    \subref{fig:fe_netsection1} cases A and D and
    \subref{fig:fe_netsection2} for cases B and C. These results are
    for a lattice of $\bar{\rho}=0.11$.}
\label{fig:damagemode}
\end{figure}

The transition flaw size is a useful parameter for quantifying the
flaw sensitivity of a lattice. It is the minimum semi-length of a
centre-crack for which catastrophic failure is dictated by the
fracture toughness $K_{\rm IC}$ of the lattice rather than the tensile
strength of the uncracked lattice. The notion of a transition flaw
size was explored for the case of a brittle lattice by Fleck and
co-workers \cite{Fleck2007,Quintana-Alonso2010,Seiler2019a} and for
the case of a ductile lattice by \citet{Tankasala2017}. In all cases,
the transition flaw size $a_{\rm T}$ is given by
\begin{equation}
  \label{eq:a_T}
  a_{\rm T} \approx \frac{1}{\pi} \left( \frac{K_{\rm IC}}{\sigma_{\rm f}^\infty}\right)^2 
\end{equation}
where $K_{\rm IC}$ is the macroscopic mode I fracture toughness of the
lattice, and ${\sigma_{\rm f}^\infty}$ is the macroscopic tensile
strength for a brittle lattice, or the macroscopic yield strength for
a ductile lattice. The transition flaw size is on the order of one
unit cell dimension ($a_{\rm T} \approx \ell$) for a brittle hexagonal
lattice for which each strut fails by bending. In contrast, the struts
of a ductile hexagonal lattice fail by stretching, and consequently,
$a_{\rm T} \approx 250\ell$, following the predictions of
\citet{Tankasala2017} for $\varepsilon_\mathrm{fs}=1.3$ in the absence
of cell wall strain hardening. Four classes of lattice behaviour can
be identified, depending upon the values of $a_{\rm T}/\ell$ and
${\varepsilon_{\rm fs,\rms}}/\bar{\varepsilon}_{\rm fs}$, see
\cref{fig:damage_mode}:
\begin{itemize}
\item[(i)] \textit{Case A} is a brittle hexagonal lattice with a
  deterministic value of strut tensile strength (or ductility) such
  that ${\varepsilon_{\rm fs,\rms}}=0$. Cell walls within the lattice
  fail at a local bending strain of only a few percent, giving rise to
  $a_{\rm T} \approx \ell$. PMMA lattices at room temperature can
  behave in this manner, see for example \citet{Seiler2019a}.
\item[(ii)] \textit{Case B} is a ductile hexagonal lattice with a
  deterministic value of strut ductility
  (${\varepsilon_{\rm fs,\rms}}=0$). When failure of a cell wall is
  dictated by the average tensile strain across a section of the
  strut, then the transition flaw size of the lattice is on the order
  of few hundreds of cells, $a_{\rm T} \approx 250\ell$ for a
  perfectly plastic cell wall solid of $\varepsilon_{\rm fs}=1.3$ as
  previously reported by \citet{Tankasala2017}.
\item[(iii)] \textit{Case C} is a ductile hexagonal lattice with a
  dispersion in strut ductility such that
  ${\varepsilon_{\rm fs,\rms}}/\bar{\varepsilon}_{\rm
    fs}=0.4$. Similar to case B, the transition flaw size is of order
  $a_{\rm T} \approx 250\ell$ when the average tensile strain across a
  section of the strut dictates strut failure. The PMMA lattices of
  the current study (tested at 100$^{\rm o}\rm C$) belong to this
  category.
\item[(iv)] \textit{Case D} is a brittle hexagonal lattice with a
  dispersion in strut tensile strength (or ductility) such that
  ${\varepsilon_{\rm fs,\rms}}/\bar{\varepsilon}_{\rm fs}=0.4$. PMMA
  lattices at room temperature can also be of this class, and possess
  a small $a_{\rm T} \approx \ell$, as confirmed recently by
  \citet{Seiler2019a}.
\end{itemize}

The above cases A through D are summarised in
\cref{fig:damage_mode}.  \citet{Seiler2019a} have shown by a
combination of experiment and FE analysis on PMMA lattices at room
temperature that the sequence of strut failure in brittle lattices of
type A and D is correlated, as sketched in
\cref{fig:damage_mode_crack}. When the failure strain of a strut is
deterministic (case A), the most highly stressed strut fails first,
and then the adjacent struts fail in a sequential, correlated
manner. A dispersion in the strut ductility of the lattice (case D)
leads to failure of struts at random locations within the lattice, but
a small transition flaw size $a_{\rm T} \approx \ell$ again leads to a
correlated failure mode of struts adjacent to the weakest one.

In contrast, the transition flaw size $a_{\rm T}$ is large compared to
the strut length of PMMA lattices at high temperature
(100$^{\rm o}\rm C$), and failure is either correlated (case B) or
diffuse (case C) depending upon the value of
${\varepsilon_{\rm fs,\rms}}/\bar{\varepsilon}_{\rm fs}$. Consider
first the case of a deterministic strut failure strain,
${\varepsilon_{\rm fs,\rms}}=0$, case B. A crack-like damage zone
develops such that failure of most highly stressed strut triggers the
creep failure of adjacent struts: a correlated damage zone develops.
Alternatively, a high value of
${\varepsilon_{\rm fs,\rms}}/\bar{\varepsilon}_{\rm fs}$, such as
${\varepsilon_{\rm fs,\rms}}/\bar{\varepsilon}_{\rm fs}=0.4$, leads to
a diffuse mode of damage. Final fracture occurs when the size of a
critical cluster of failed struts exceeds the least of $2a_{\rm T}$
and $0.5W$; this is shown in case C of \cref{fig:damage_mode_crack}.
The approximate position of the boundary between diffuse and correlated modes of damage is shown in \cref{fig:damage_mode} based on the available data. A much more detailed and exhaustive study is needed to give the precise location of this boundary.

The damage tolerance of a lattice can be assessed, alternatively, by
plotting the normalised macroscopic stress $\sigma^{\infty}$ versus
number of failed struts, $n$, across the width of the lattice. Here,
$\bar{\sigma}_{\rm f}^{\infty}$ is the mean macroscopic strength for
first strut failure from 5 FE simulations and 3 experiments. Cases A
through D are plotted in \cref{fig:fe_netsection1,fig:fe_netsection2}
from both the experiments and FE. The results for cases A and D are
taken from Ref.~\cite{Seiler2019a}, whereas cases B and C are from the
present study. Recall that case C denotes a diffuse mode of strut
failure and a large value of $a_{\rm T}$. The macroscopic stress
$\sigma^{\infty}$ drops slightly with an increasing number $n$ of
failed struts, see \cref{fig:fe_netsection2}.  In contrast, the drop
in $\sigma^{\infty}$ with increasing $n$ follows the net section
prediction,
$\sigma^{\infty}/\bar{\sigma}_{\rm f}^{\infty}=1-(n-1)/(N-1)$, for
case B of deterministic strut ductility
$({\varepsilon_{\rm fs,\rms}}/\bar{\varepsilon}_{\rm fs}=0)$. Here $N$
is the total number of struts in the gauge section; $N$ equals $12$ in
\cref{fig:fe_netsection1,fig:fe_netsection2}. This suggests that a
dispersion in strut ductility is beneficial for the enhancement of
damage tolerance in the creep regime.

\section{The effect of as-designed defects on macroscopic properties:
  experiment}
\label{sec:asdesigned_defects}
Three types of macroscopic defect were introduced within the regular
lattice by design: (i) misplaced joints, (ii) cells filled with solid
inclusions, and (iii) missing cell walls. The resulting
as-manufactured specimens contain geometric and material imperfections
at the cell wall level (variation in strut thickness, in Plateau
border radius, and in strut ductility) in addition to one of the three
macroscopic defects, see \cref{fig:irrHexLattice}. The macroscopic
tensile strength at first strut failure of the as-manufactured
specimens was measured experimentally at 100$^{\rm o}\rm C$ and a
strain rate of \SI{4E-4}{s^{-1}} and then compared with that of the
as-manufactured topologies designed without macroscopic defects. The
sensitivity of measured creep strength to the presence of as-designed
defects was thereby assessed.

\subsection{Randomly perturbed joints}
\label{sec:perturbations_nodal}
The effect of cell wall misalignment on the macroscopic creep response
of PMMA lattices was investigated by generating imperfect lattice
topologies with randomly perturbed joints. These lattices were
laser-cut based on a CAD file wherein the joints of a regular
hexagonal lattice (of $\bar\rho=0.11$) were repositioned randomly
within a circular disc of radius $R$, following the procedure of
\citet{Romijn2007}. The degree of imperfection was varied by selecting
values of $R/\ell$ between $0$ (regular lattice) and $0.5$ (extremely
imperfect lattice). A typical realisation of the as-manufactured
lattice, for the choice $R/\ell=0.5$, is shown in
\cref{fig:hexLattice_2}; only those joints which lie within the gauge
section were misplaced. The random misplacement of the joints reduces
the average strut length such that the relative density of the lattice
increases by a factor of 1.0025 for $R/\ell=0.1$ and by a factor of
1.0625 for $R/\ell=0.5$, as previously noted by \citet{Romijn2007}.
This minor change in $\bar\rho$ is ignored in the current study.

\begin{figure}[htbp]
  \centering \setlength{\labelsep}{-1cm}
  \sidesubfloat[]{\includegraphics[width=0.49\textwidth]{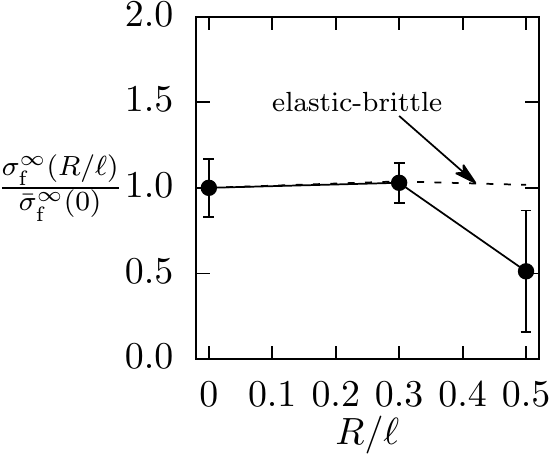}\label{fig:misplacedNodes_stress_HT_statistic}}\hfill
  \sidesubfloat[]{\includegraphics[width=0.49\textwidth]{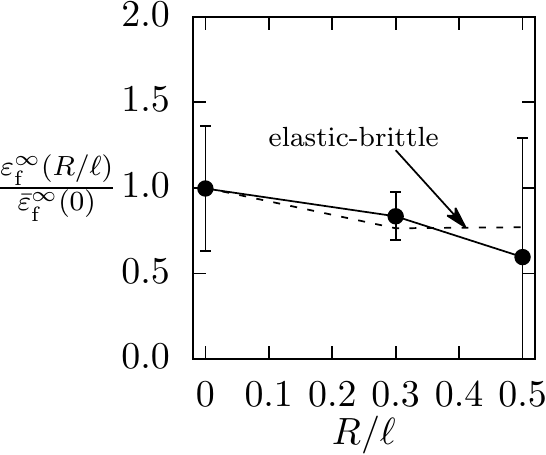}\label{fig:misplacedNodes_strain_HT_statistic}}\\[\baselineskip]
  \setlength{\labelsep}{0.2cm}
  \sidesubfloat[]{\includegraphics[width=0.48\textwidth]{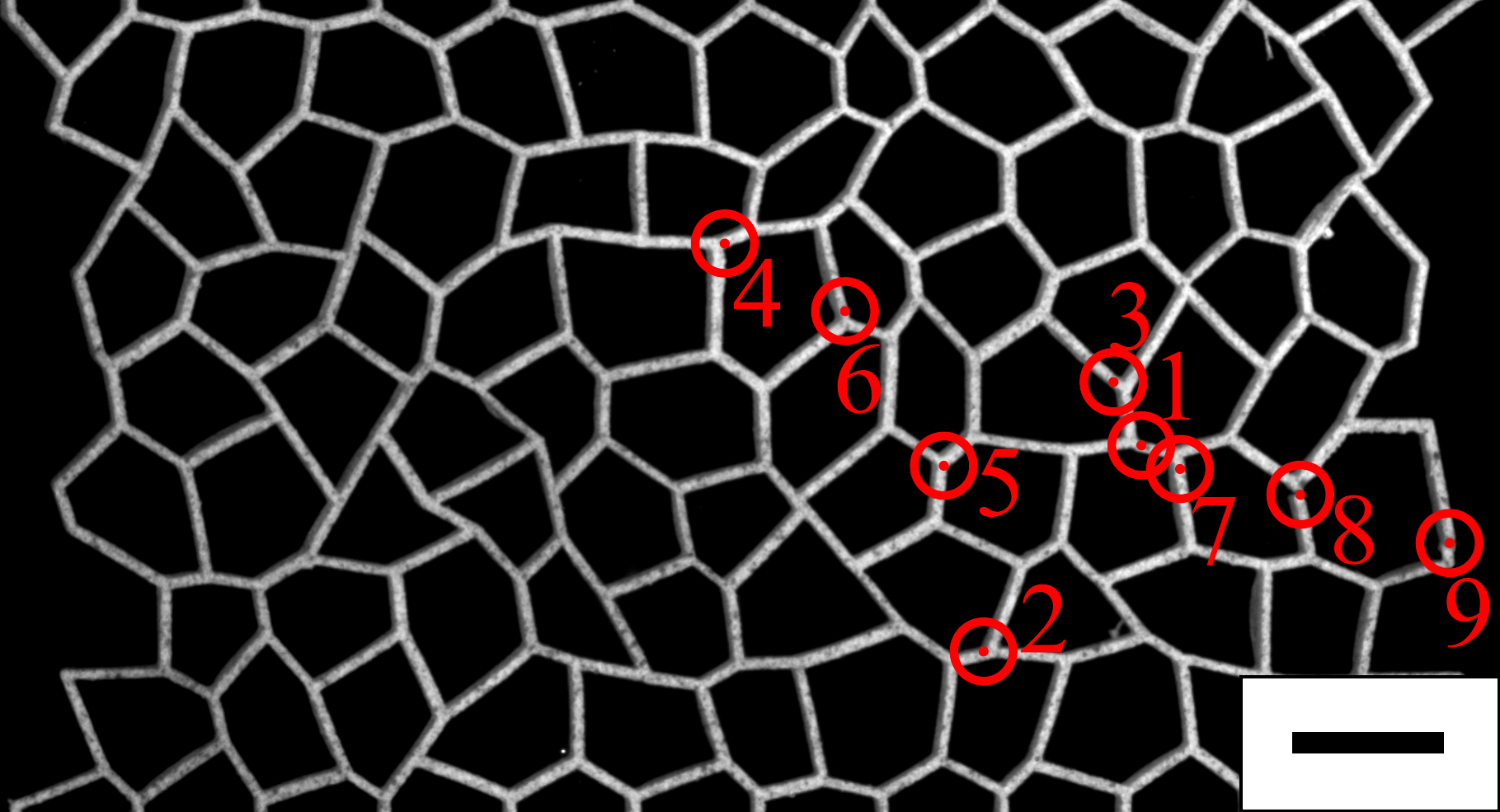}\label{fig:hexlattice_short_cut_100C_14_failure}}
  \caption{Measured macroscopic properties of imperfect lattices with
    randomly misplaced joints:
    \subref{fig:misplacedNodes_stress_HT_statistic} macroscopic
    tensile strength $\sigma^\infty_\mathrm{f}$ and
    \subref{fig:misplacedNodes_strain_HT_statistic} macroscopic
    tensile ductility $\varepsilon^\infty_\mathrm{f}$ for first strut
    failure. The corresponding measured responses for elastic-brittle
    PMMA lattices as taken from \citet{Seiler2019a} are included for
    comparison.  \subref{fig:hexlattice_short_cut_100C_14_failure}
    Sequence of strut failure in one specimen of ($R/\ell=0.5$). The
    scale bar is of length 10 mm.}
   \label{fig:perturbed_joints}
\end{figure}

The sensitivity of macroscopic tensile strength
$\sigma^\infty_\mathrm{f}$ at first strut failure, and macroscopic
tensile ductility $\varepsilon^\infty_\mathrm{f}$ at first strut
failure, to random perturbation of joints was measured for 3
structural realisations for each choice of $R/\ell=0, 0.3,$ and $0.5$,
all for relative density $\bar\rho=0.11$. The measured values of
$\sigma^\infty_\mathrm{f}$ and $\varepsilon^\infty_\mathrm{f}$ are
plotted in \cref{fig:perturbed_joints} as a function of the degree of
imperfection $R/\ell$. The ordinate in each case is normalised by its
corresponding mean value as measured for the regular lattice
($R/\ell=0$). For comparison, the knock-down in
$\sigma^\infty_\mathrm{f}$ with increasing $R/\ell$ for the case of
elastic-brittle PMMA lattices (of $\bar\rho=0.11$) is included in
\cref{fig:perturbed_joints}. The mean values of
$\sigma^\infty_\mathrm{f}$ are obtained from 3 realisations.

No significant knock-down in macroscopic tensile strength and
macroscopic tensile ductility was found for $0 < R/\ell \leq
0.3$. Significant scatter in both $\sigma^\infty_\mathrm{f}$ and
$\varepsilon^\infty_\mathrm{f}$ is observed when $R/\ell$ is increased
from 0.3 to 0.5. The mean values of $\sigma^\infty_\mathrm{f}$ and
$\varepsilon^\infty_\mathrm{f}$ are reduced by a factor of 2 and 1.6,
respectively, from $R/\ell=0.3$ to $R/\ell=0.5$. This knock-down in
$\sigma^\infty_\mathrm{f}$ and $\varepsilon^\infty_\mathrm{f}$ is
attributed to the development of discrete force chains spanning the
length of the specimen, wherein the struts deform predominantly by
stretching. This observation is consistent with the FE predictions of
\citet{Tankasala2017} which revealed that few force chains in the
lattice carry most of the applied load, and the stretching-dominated
failure of these struts results in a reduction of the macroscopic
ductility. The sequence of strut failure in a representative specimen
of $R/\ell=0.5$ in \cref{fig:hexlattice_short_cut_100C_14_failure}
reveals a diffuse mode of damage, similar to the $R/\ell=0$ case. For
all realisations of $R/\ell>0$ tested in this study, first strut to
fail is always located at the centre of the specimen.

\subsection{Missing cell walls}
\label{sec:missing_cell_walls}
A centre-crack was introduced in regular lattice specimens by removing
a row of cell walls: the cell walls comprising the crack were first heated
by a soldering iron. They were then cut whilst the deformation response of
the PMMA strut was in the rubbery regime, to obtain specimen geometry
as sketched in \cref{fig:hexLattice_3}. The initial crack is of
semi-length $a_0=\sqrt{3} n_\mathrm{b} \ell /2$ where $n_{\rm b}$ is
the number of broken cell walls. The influence of crack length on the
tensile strength of the lattice was explored by varying $n_{\rm b}$
between 0 and 6; 3 realisations of the lattice were generated for each
value of $n_\mathrm{b}$, with $\bar\rho=0.11$.

\begin{figure}[htbp]
  \centering \setlength{\labelsep}{-1cm}
  \sidesubfloat[]{\includegraphics[width=0.49\textwidth]{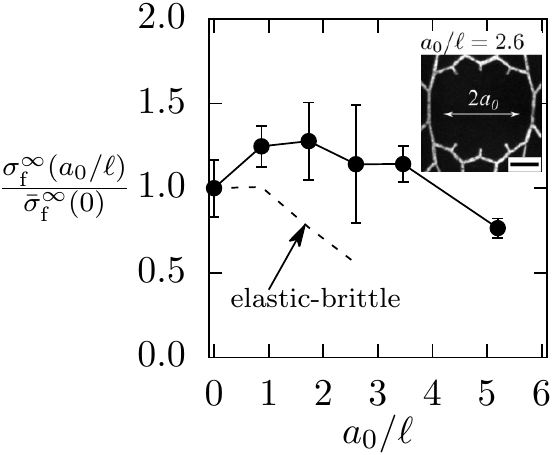}\label{fig:removedCellWalls_stress_HT_statistic}}\hfill
  \sidesubfloat[]{\includegraphics[width=0.49\textwidth]{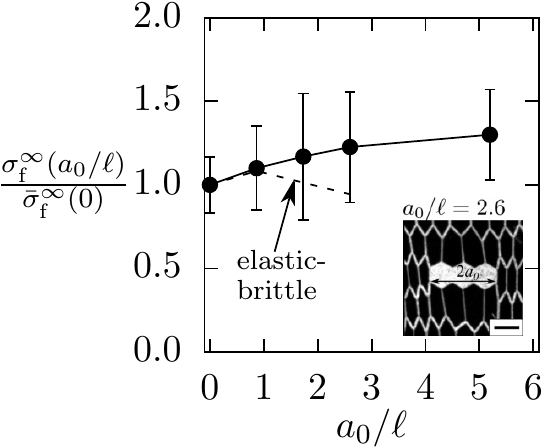}\label{fig:filledCells_stress_HT_statistic}}\\[\baselineskip]
  \setlength{\labelsep}{0.2cm}
  \sidesubfloat[]{\includegraphics[width=0.49\textwidth]{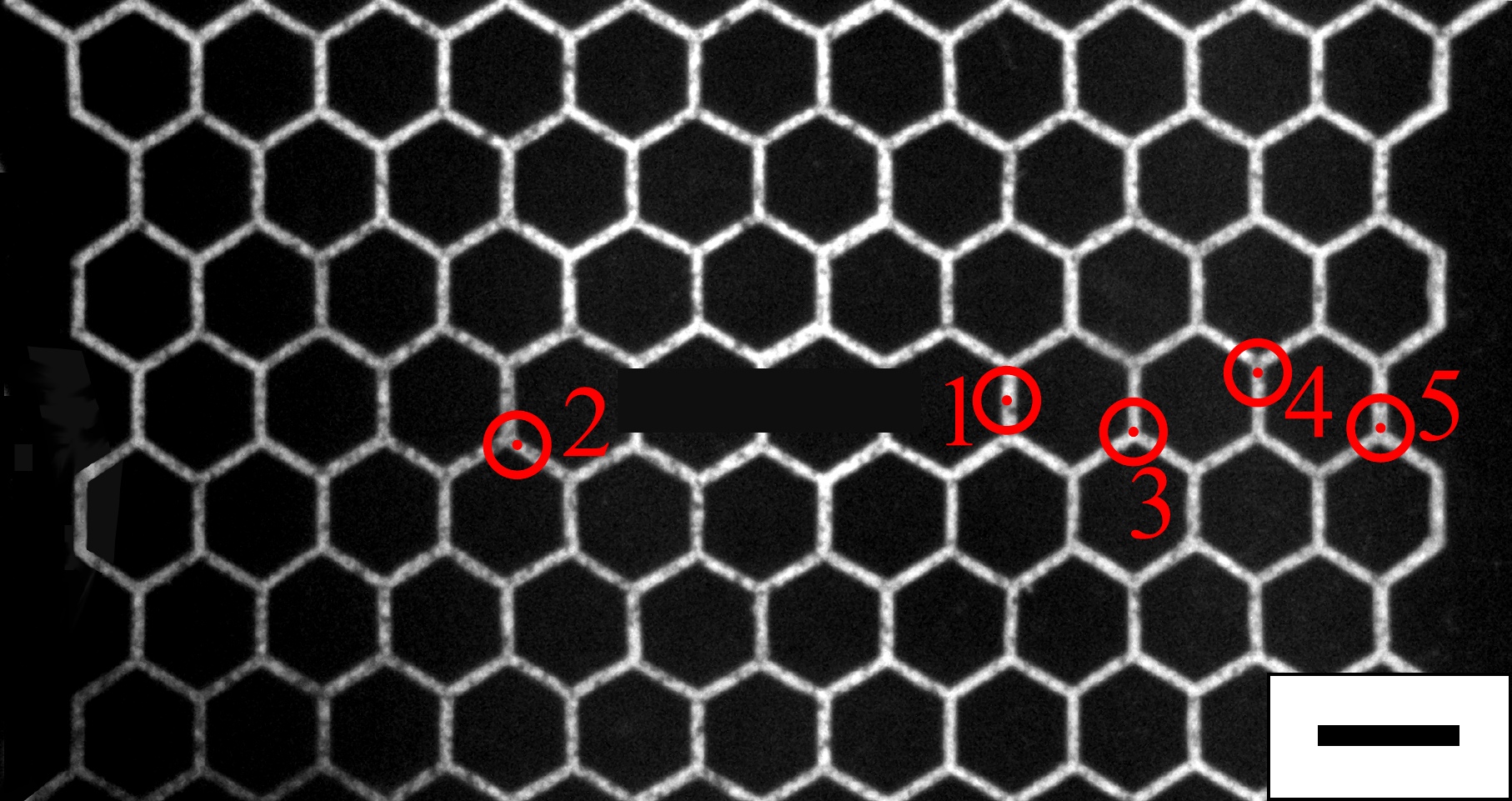}\label{fig:hexlattice_short_cut_100C_5_failure}}\\[\baselineskip]
  \sidesubfloat[]{\includegraphics[width=0.49\textwidth]{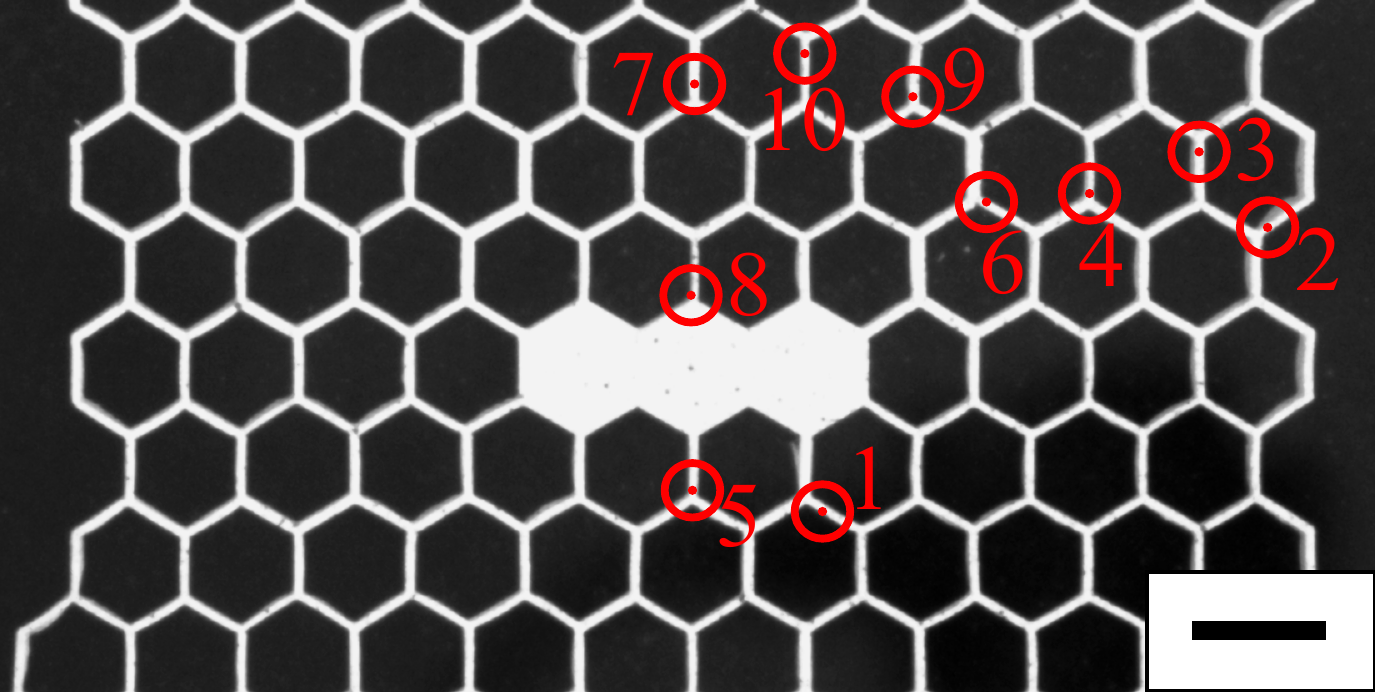}\label{fig:hexlattice_short_cut_100C_43}}
  \caption{Measured tensile strength of lattice specimens with
    \subref{fig:removedCellWalls_stress_HT_statistic} a row of missing
    cell walls and \subref{fig:filledCells_stress_HT_statistic} a row
    of solid inclusions, as a function of the defect size
    $a_0/\ell$. The corresponding measured responses for
    elastic-brittle PMMA lattices as taken from \citet{Seiler2019a}
    are included for comparison. A magnified view of the deformed
    lattice just prior to first strut failure is shown in the
    insets. Sequence of strut failure in one specimen containing
    \subref{fig:hexlattice_short_cut_100C_5_failure} missing cell
    walls ($a_0/\ell = 2.6$) or
    \subref{fig:hexlattice_short_cut_100C_43} solid inclusion
    ($a_0/\ell = 2.6$). The scale bars are of length 10 mm.}
   \label{fig:imperfect_lattices}
\end{figure}

The sensitivity of macroscopic tensile strength
$\sigma^\infty_\mathrm{f}$ at first strut failure to the presence of a
macroscopic crack is shown in
\cref{fig:removedCellWalls_stress_HT_statistic} as a function of the
crack length $a_0/\ell$.  As before, the ordinate in
\cref{fig:removedCellWalls_stress_HT_statistic} is normalised by its
corresponding mean value as measured for the uncracked lattice
($a_0/\ell=0$). The ductile hexagonal lattice has a high transition flaw
size $a_{\rm T}$ such that only a small knock-down in tensile strength
is observed for $a_0/l \leq 5$. This behaviour is in contrast to the
response of centre-cracked brittle PMMA lattices in which the tensile
strength was observed to be governed by the fracture toughness of the
lattice, see~\cite{Seiler2019a} for details. \changes{Note that the
  net section strength of the ductile lattice always lies above the gross section strength,
  and so the measured strengths of
  \cref{fig:removedCellWalls_stress_HT_statistic} indicate a
  significant elevation in net section strength at small (but finite)
  $a_0/\ell$. The reason for this strengthening is unclear, and is a
  topic for future research. The available evidence in the insert of
  \cref{fig:removedCellWalls_stress_HT_statistic} suggests a blunting
  mechanism at the tip of the pre-crack, whereby the struts align with
  the loading direction. This crack tip blunting phenomenon has been
  analysed by \citet{Tankasala2015} for the case of a long crack in a ductile 
  hexagonal lattice. This explanation is also consistent with that of
  \citet{Mangipudi2011}. They calculated the net section strength of a
  centre-cracked panel (CCP) of a honeycomb of random Voronoi
  microstructure. They modelled the failure of each strut by using
  softening beam elements, and found a small elevation in net section
  strength when the size of the pre-crack was increased from $a/W=0$ to
  $a/W = 0.1$. \citet{Mangipudi2011} noted that the strengthening is associated with a
  change in fracture path from a random, uncorrelated manner to failure originating from
  the crack tip. This is consistent with the observed switch in
  failure path in the present study, compare \cref{fig:Sample_0039}
  for $a_0/\ell = 0$ with
  \cref{fig:hexlattice_short_cut_100C_5_failure} for
  $a_0/\ell = 2.6$.} For initial cracks
of semi-length $a_0/\ell \ge 2.6$, first strut failure occurs near the
tip of the macroscopic crack, as shown in 
\cref{fig:hexlattice_short_cut_100C_5_failure}. Subsequent strut
failure occurs in a correlated manner with the advancing crack
predominantly comprising of vertical struts. 

\subsection{Solid inclusions}
\label{sec:solid_inclusions}
Hexagonal lattices containing a solid inclusion were generated by the
laser-cutting of PMMA sheets, with a number of intact filled cells at
the centre of the specimen, recall \cref{fig:hexLattice_4}. The
semi-length of the inclusion is $a_0=\sqrt{3}n_{\rm c}\ell/2$ where
$n_{\rm c}$ is the number of filled cells. Three realisations of the
lattice were generated for each value of $n_{\rm c}$ between 0 and 6,
with $\bar{\rho}=0.11$.

The measured values of macroscopic tensile strength
$\sigma^\infty_\mathrm{f}$ at first strut failure are plotted in
\cref{fig:filledCells_stress_HT_statistic} as a function of the
inclusion size $a_0/\ell$. The sensitivity of
$\sigma^\infty_\mathrm{f}$ to $a_0/\ell$ for the case of
elastic-brittle lattice as taken from Ref.~\cite{Seiler2019a} are
included in \cref{fig:filledCells_stress_HT_statistic} for
comparison. A mild increase in $\sigma^\infty_\mathrm{f}$ is observed
for increasing values of $a_0/\ell$. This is due to a reduction in the
lateral contraction of cells in the vicinity of the
inclusions. Diffuse strut failure is observed prior to catastrophic
fracture with the failed struts not necessarily located in the
vicinity of the inclusions, see for example the case of $a_0/\ell=2.6$
in \cref{fig:hexlattice_short_cut_100C_43}.

\section{Concluding remarks}
\label{sec:conclusin}
The present study explores the sensitivity of the macroscopic tensile
response of a visco-plastic lattice to as-manufactured and as-designed
defects. It is found that both classes of defect have a significant
effect on the macroscopic tensile strength of laser-cut PMMA lattices
in the visco-plastic regime. FE analysis provides further insight into
the relative potency of three types of as-manufactured defect: a
dispersion in strut thickness, in Plateau border radius and in strut
ductility arising from the manufacturing route.

The evolution of strut failure at $T$ close to $T_\mathrm{g}$ is
markedly different from that observed at $T<T_\mathrm{g}$ in a
companion study, Ref. \cite{Seiler2019a}. At room temperature, PMMA behaves
in an elastic-brittle manner and the failure of a single strut is
sufficient to lead to the formation of a single macroscopic crack in
the lattice. In contrast, at $T \approx T_\mathrm{g}$, a diffuse zone
of failed struts develop prior to catastrophic failure. Finite element
analysis reveals that a dispersion in strut-to-strut ductility also
plays a major role in giving rise to the diffuse damage state in the
creep regime. A dispersion in ductility from strut to strut is
essential to give rise to the observed early onset of strut failure as
well as the diffuse mode of fracture. The latter in turn enhances the damage tolerance of these lattices in creep regime. In contrast, a dispersion in strut thickness and Plateau border radius have only a mild effect in the
creep regime, as confirmed by the FE predictions.

As-designed imperfections were also introduced in the laser-cut
lattices. Three kinds of imperfections were explored experimentally:
randomly misplaced joints, a row of missing cell walls to create a
notch, and a row of filled cells as solid inclusions. The following
conclusions can be drawn for each type of defect:

\begin{itemize}
\item[(i)] \textit{Imperfections in the form of randomly perturbed
    joints.}  The macroscopic tensile strength of the lattice is
  insensitive to random perturbation of joints for low values of
  $R/\ell$ between 0 and 0.3. A 50\% knock-down in tensile strength is
  observed when the imperfection is most severe ($R/\ell=0.5$). The
  macroscopic ductility decreases with increasing values of $R/\ell$
  due to the formation of discrete force chains. This behaviour is in
  contrast to brittle hexagonal lattices which are almost insensitive
  to the presence of randomly perturbed joints.

\item[(ii)] \textit{Imperfections in the form of missing cell walls.}
  Visco-plastic lattices possess high damage tolerance. The transition
  flaw size $a_\mathrm{T}$ for strength-controlled failure to fracture
  toughness-controlled failure is sufficiently large such that the
  macroscopic tensile strength is not reduced by the presence of even
  4 missing cells. This observation is consistent with the numerical
  predictions of \citet{Tankasala2017} for ductile hexagonal lattices.
  
\item[(iii)] \textit{Imperfections in the form of solid inclusions.}
  Lattices with filled cells show a slight increase in tensile
  strength due to the reduced lateral contraction of cells in the
  vicinity of the inclusions.
\end{itemize}

In summary, the macroscopic tensile strength of the visco-plastic
hexagonal PMMA lattice is almost insensitive to imperfections in the
form of broken cell walls and solid inclusions owing to its high
transition flaw size. The random misplacement of joints emerges as the
most potent type of defect.

\section*{Acknowledgements}
\noindent The authors gratefully acknowledge the financial support
from the European Research Council (ERC) under the European Union's
Horizon 2020 research and innovation program, grant GA669764,
MULTILAT.

\section*{References}
\bibliographystyle{elsarticle-num-names}
\bibliography{LiteratureCambridge}

\newpage
\newpage
\section*{Figures}
\floatsetup[figure]{subcapbesideposition=top}

\end{document}